\documentclass[aps,twocolumn,groupedaddress,superscriptaddress]{revtex4-1}
\usepackage{graphicx}
\usepackage{amsmath,amssymb}
\usepackage{textcomp}
\usepackage{float}
\begin{document}

\begin{titlepage}
\title{Force networks and jamming in shear deformed sphere packings}
\author{H. A. Vinutha}
\affiliation{Jawaharlal Nehru Center for Advanced Scientific Research, Jakkur Campus, Bengaluru 560064, India.}
\affiliation{TIFR Center for Interdisciplinary Sciences, 21 Brundavan Colony, Narsingi, Hyderabad  500075, India.}
\author{Srikanth Sastry}
\affiliation{Jawaharlal Nehru Center for Advanced Scientific Research, Jakkur Campus, Bengaluru 560064, India.}

\begin{abstract}
{The formation of self-organised structures that resist shear deformation have been discussed in the context of shear jamming and thickening\cite{bi2011,mari2015,peters2016}, with frictional forces playing a key role. However, shear induces geometric features necessary for jamming even in frictionless packings\cite{vinu2016}. We analyse conditions for jamming in such assemblies  by solving force and torque balance conditions for their contact geometry. We demonstrate, and validate with frictional simulations, that the  {\textrm {\textit{isostatic condition}}} for mean contact number $Z = D + 1$ (for spatial dimension $D = 2$, $3$) holds at jamming for both finite and infinite friction, above the {\textrm {\textit{random loose packing}}} density. We show that the shear jamming threshold satisfies the marginal stability condition recently proposed for jamming in frictionless systems\cite{wyart2012}. We perform rigidity percolation analysis\cite{thorpe1983,henkes2016} for $D = 2$ and  find that rigidity percolation precedes shear jamming, which however coincides with the percolation of over-constrained regions, leading to  the identification of an {\textrm {\textit{intermediate phase}}} analogous to that observed in covalent glasses\cite{intermediate}. }
\end{abstract}

\maketitle
\end{titlepage}


Jamming is the process by which disordered assemblies of particles become rigid and resist externally imposed stresses, for instance when their density becomes large enough.  It has been widely investigated, both as a phenomenon that occurs in granular matter, and as a particular aspect of the emergence of rigidity in disordered matter, {\it e. g.} colloidal suspensions, foams, glass formers and gels and to understand the rheological properties of thermal and athermal driven systems \cite{liu-2010,zhang2010,bi2011,wyart2012,mari2015,peters2016}.
The jamming of frictionless sphere assemblies is particularly well studied and occurs at a packing fraction of $\phi_J \approx 0.64$, referred to as  random close packing (RCP) or the jamming point. In the presence of friction, jamming is expected to occur down to a significantly lower density, which is $\sim 0.54$ (in $3D$) \cite{silbert2010,makse2008} in the isotropic case, also known as the random loose packing density (RLP), but strong dependences on friction and protocol lead to a wide range of estimates of this density,  [$0.54 - 0.61$].  A rather different scenario was envisaged by Cates {\it et al}  \cite{cates1998} for jamming in systems subjected to external stress, in which the application of external stress itself leads to a self organisation of particles that could resist stress, and thus lead to jamming. Such a scenario of {\it shear jamming} has been studied recently experimentally  and theoretically \cite{zhang2010,bi2011,sarkar2013,sarkar2016} for sheared granular packings in the presence of friction, but also in the case of frictionless spheres \cite{bertrand2016,kumar2016,baity2016,urbani-2017}. However, our understanding is yet incomplete concerning various central issues, such as: 
(i) the range of densities over which shear jamming may occur, and the corresponding conditions, (ii) the differences and similarities between shear jamming and the isotropic frictionless as well as frictional jamming, (iii) a geometric description of the self organization of particles that lead to jamming behaviour, and (iv) the origins of the geometric organisation observed.\\

We address these issues in the present work, by exploiting the observation \cite{vinu2016} that athermally sheared frictionless sphere assemblies develop structural features that correspond to shear jamming, when frictional forces are present to provide mechanical stability. Thus, the geometry of such assemblies, encapsulated in their contact network can be interrogated to understand geometric and mechanical conditions necessary for shear jamming, avoiding some of the ambiguities attendant simulation and experimental investigations of frictional packings. We do so by solving force and torque balance conditions that must be satisfied, with finite contact forces, by jammed states, for both finite and infinite friction. We employ a new method which improves the accuracy of such computations, and validate our results through simulations of frictional particles.  Details of our computations are provided in the Methods section and in the Suppmentary Information (SI). As described below, we obtain a precise, detailed geometric characterisation of the shear jamming transition that is at variance in key aspects with results for isotropic frictional jamming \cite{makse2008}, and makes contact with analyses of rigidity in the rather different context of covalent network glasses \cite{intermediate}. We also obtain a mechanical characterisation, that is consistent with criteria for marginal stability analysed for frictionless packings but not hitherto applied to shear or frictional jamming.


\begin{figure}[htpb!]
\hspace*{-0.5cm}
\includegraphics[scale=0.3]{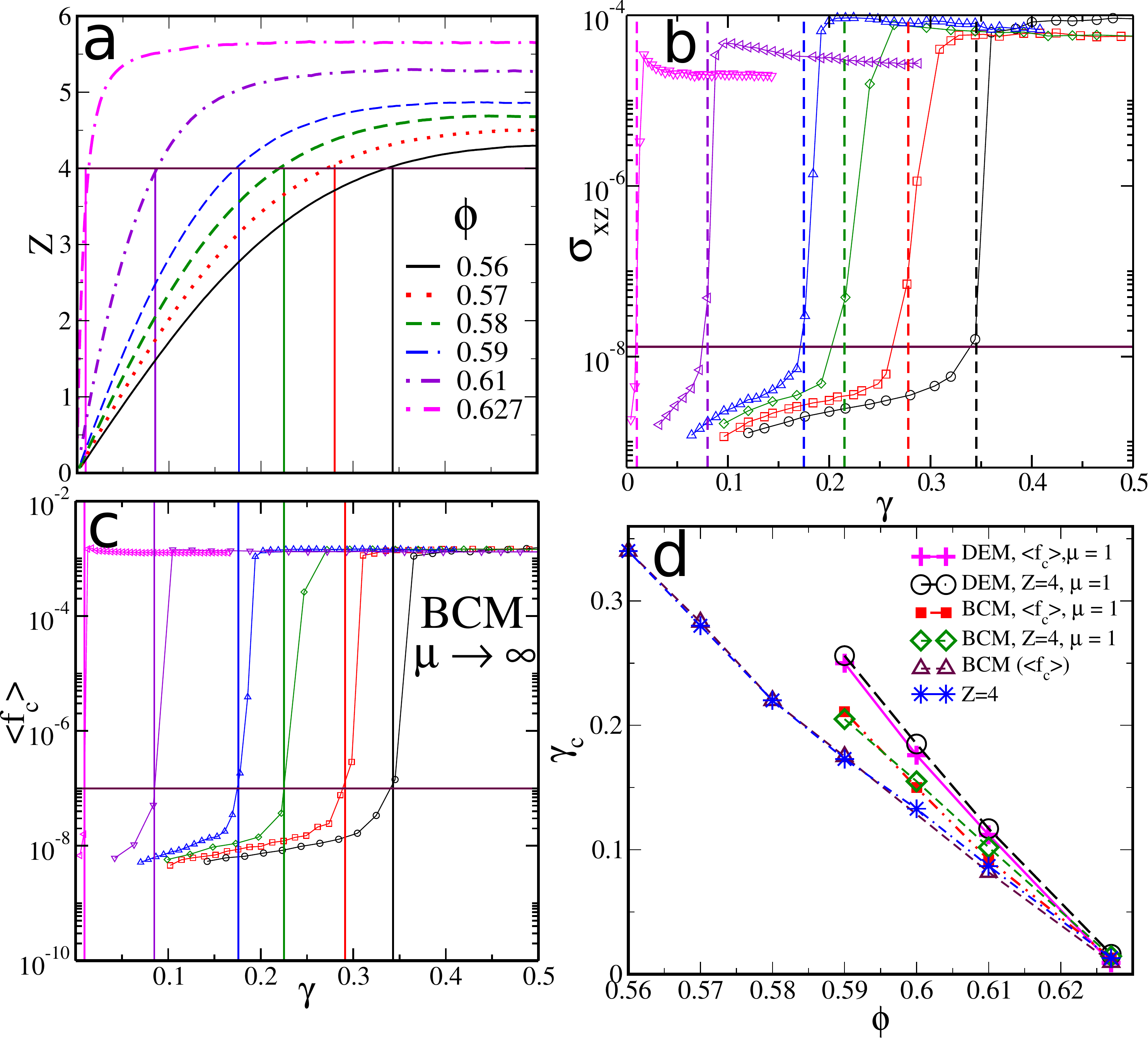}
\caption{{\bf (a)} The average contact number $Z$, {\bf (b)} stress $\sigma_{xz}$, and {\bf (c)} the average contact force $\langle f_c \rangle$, as a function of strain, shown for different densities. $Z$ values are obtained from sheared configurations, and $\sigma_{xz}$ and $f_c$ from force balance solutions.  The vertical lines show strain values corresponding to the shear jamming (SJ) transition, where $\sigma_{xz}$ and $\langle f_c \rangle$ show discontinuous jumps, which also correspond to $Z = 4  (= D+1)$. The maroon horizontal lines indicates $Z=4$ and the cutoff used to identify the shear jamming  strain in $\sigma_{xz}$ and $\langle f_c \rangle$ plots. {\bf (d)} Threshold shear strains as a function of density, which shows that the shear jamming (SJ) transition, in the limit of infinite friction and finite friction, occurs at $Z=D+1$.}
\label{figm1}  
\end{figure}
\begin{figure}[h!]
\hspace*{-0.6cm}
\includegraphics[scale=0.33,angle=0]{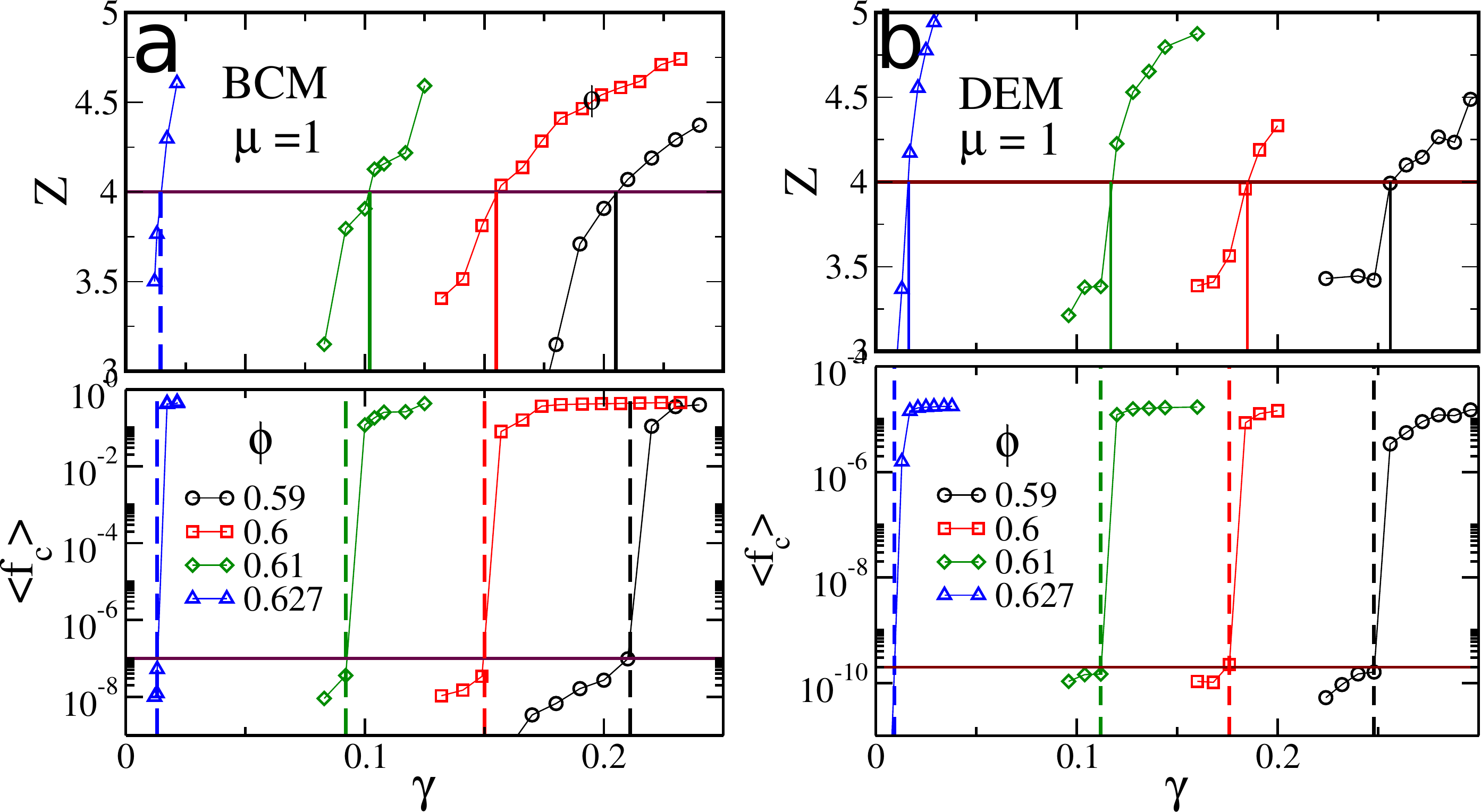}
\caption{\label{figm2} The average contact number $Z$ and the average contact force $\langle f_c \rangle$, for {\bf (a)} BCM and {\bf (b)} DEM, $\mu=1$. $Z$ values for BCM ($\mu=1$) are obtained after removing rattler contacts, {\it i. e.}, contacts with $f_c \leq 10^{-10}$. For the DEM case, only contacts that remain after the simulations are considered. The vertical (dashed) lines shows strain values corresponding to the shear jamming (SJ) transition, where $\langle f_c \rangle$ show discontinuous jumps, and the bold vertical lines show strain values corresponding  to $Z = 4 (= D+1)$. The maroon horizontal lines indicates $Z=4$ and the cutoff used to identify SJ strain in $\langle f_c \rangle$ plots.} 
\end{figure}

We first consider three dimensional sphere assemblies that are athermally sheared, and consider force and torque balance conditions as a function of strain to estimate the jamming strain, in the limit of infinite friction (friction coefficient $\mu \rightarrow \infty$). Fig. \ref{figm1}(a) shows the mean contact number $Z$ for a range of densities from $0.56$ to $0.627$.  Fig. \ref{figm1}(b) and (c) show respectively the shear stress $\sigma_{xz}$, where $xz$ is the shear plane and the average contact force $\langle f_c \rangle $ obtained from the solutions to the balance conditions, where averages are performed over contacts and over independent solutions. The shear stress and the average contact force exhibit sharp increases at density dependent strain values. We identify the shear jamming transition from the discontinuous change in $\langle f_c \rangle $, which closely corresponds to the strain values where stress shows a sharp increase.  The corresponding strain value (vertical lines) is termed the jamming strain. The jamming strain at each density closely corresponds to a strain value at which $Z=D+1$ (as shown in Fig. \ref{figm1}(a)) which is the {\it isostatic} value in the infinite friction limit, based on the constraining counting argument due to Maxwell \cite{maxwell1864}. In Fig. \ref{figm1}(d), we show the jamming strain values along with the strain values at which  $Z=D+1$.  We next consider the finite friction case, with $\mu = 1$. In Fig. \ref{figm2}(a)(b), we show $Z$ and $<f_c>$ as a function of strain for each density, obtained from solutions to the balance conditions (marked BCM $\mu=1$) and from frictional simulations (marked DEM $\mu=1$). Note that in the case of BCM, only contacts with finite contact forces (with threshold $f_c < 10^{-10}$) are counted, whereas with DEM, contacts are lost during the frictional simulations. 
As shown in Fig. \ref{figm1}(d), It is clear that for finite friction too, jamming strain values correspond to $Z=4 (= D+1)$, at all densities (however, the jamming strain values differ for BCM and DEM, a feature that may depend on protocol details of DEM and need further investigation). This result for shear jamming is at variance with arguments and results for isotropic frictional jamming, wherein $Z$ varies continuously from $6(2D)$ to $4(D+1)$ as $\mu$ is varied from $0$ to $\infty$ \cite{silbert2010,makse2008}. 

\begin{figure}[h!]
\hspace*{-0.8cm}
\includegraphics[scale=0.33,angle=0]{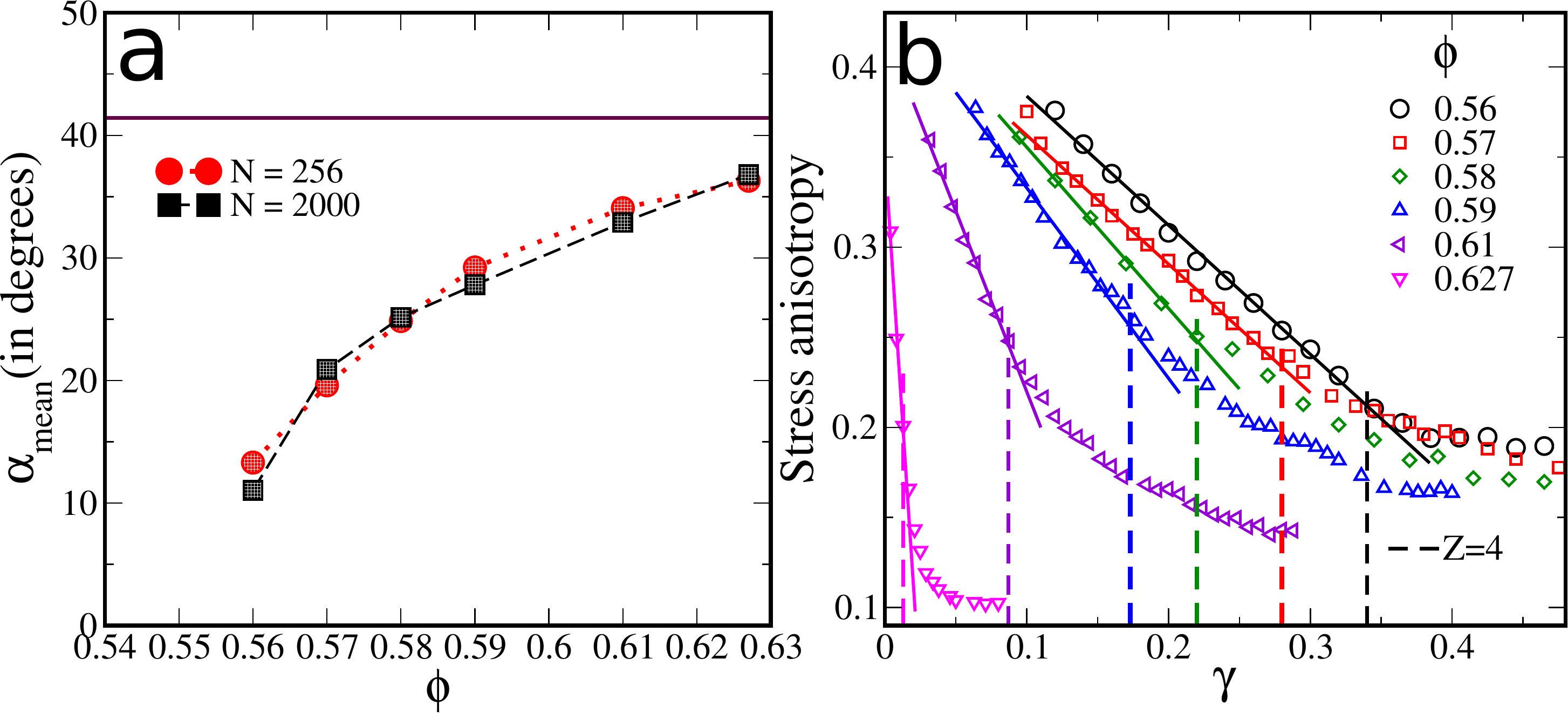}
\caption{\label{figm3} {\bf(a)} Mean angle ($\alpha_{mean}$) averaged over all pairs of independent solutions as a function of density. For comparison, we show the mean angle $\alpha_{mean} = 41.4^{\textdegree}$ (maroon horizontal line) between two vectors chosen randomly with all the components positive (first quadrant), which is greater than the mean angle obtained from force solutions at $\phi=0.627$. As the density is decreased along with the decrease in the null space dimension, see SI Fig. $S3$, $\alpha_{mean}$ mean also decreases and approaches zero at the isostatic limit.  {\bf(b)} Stress anisotropy as a function of strain for different densities, indicating saturation across the strain value where $Z$ equals $D + 1$ (indicated by dashed vertical lines). Stresses are obtained from the BCM method in the limit of infinite friction.} 
\end{figure}

We characterize the ensemble of independent force solutions by measuring the mean angle $\alpha_{mean}$ between pairs of independent solutions, In Fig. \ref{figm3}(a), we show for two different system sizes the mean angle as a function of density, for $\mu \rightarrow \infty$. $\alpha_{mean}$ decreases with a decrease in density, approaching zero as the lower limit density limit $\phi=0.55$ is approached,  indicating that decrease of the solution space volume over and above the decrease in its dimensionality (see Methods). The rapid decrease below  $\phi=0.58$ also helps explain the difficulty of finding force balanced configurations at lower densities,  as noted in previous work \cite{vinu2016,vinu-sjperc}. Another characteristic feature of forces that signals shear jamming is the saturation of the spatial anisotropy of stresses. In Fig. \ref{figm3}(b), we show the stress anisotropy (defined in Methods) as a function of strain for different densities. Initially we observe a linear increase in the stress anisotropy with strain, which  flattens out above a strain value which closely corresponds to $Z=D+1$, indicated by dashed lines for each density. 
 
\begin{figure}[]
\includegraphics[scale=0.3,angle=0]{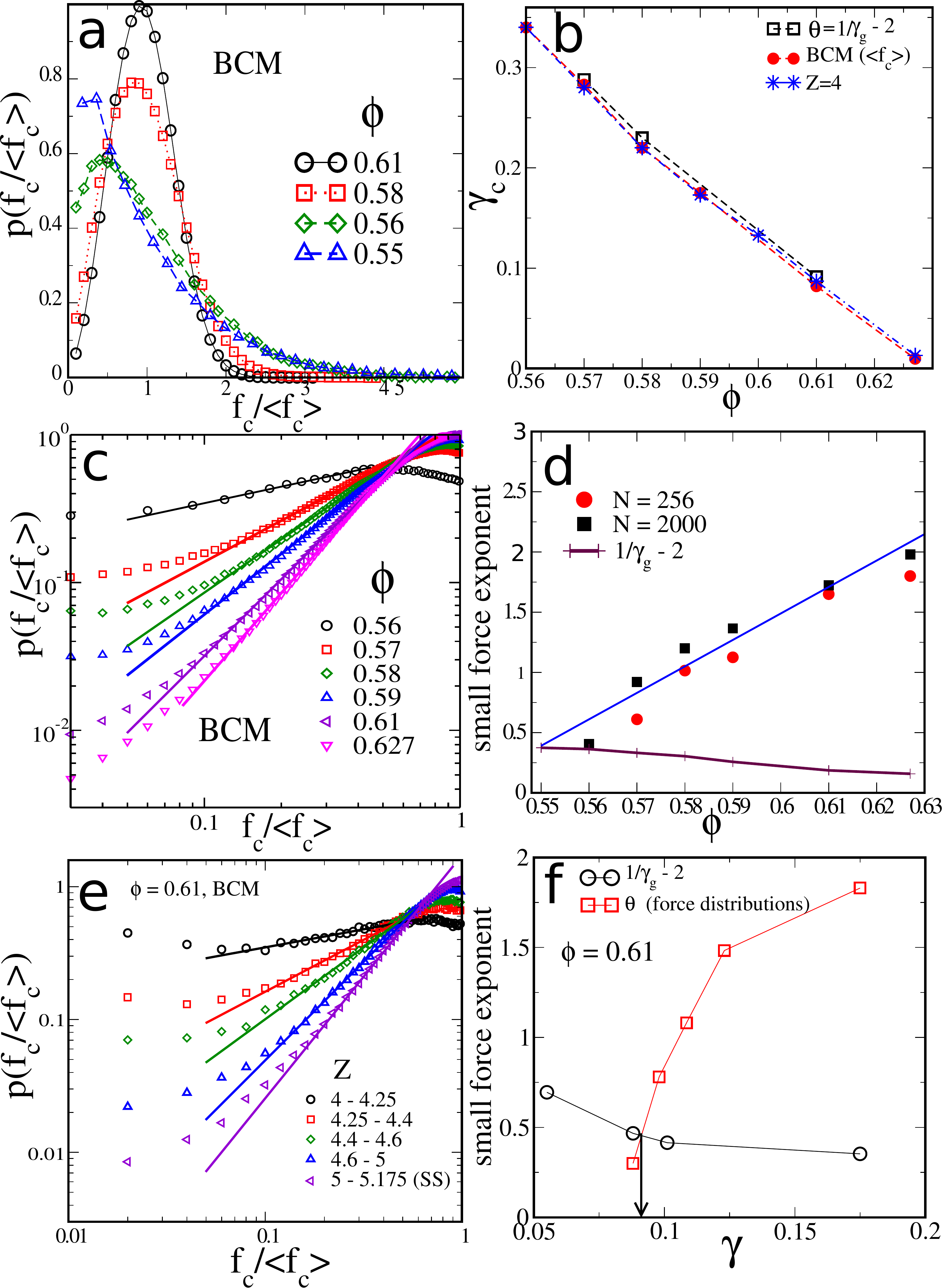}
\caption{\label{figm4} {\bf (a)}Contact force distributions obtained from BCM for the steady state configurations at different densities. Solutions are obtained starting with random initial guesses for the contact forces. Above $\phi=0.55$, the force distributions acquire jammed-like character {\it i.e.,} display finite force peaks. {\bf (b)} Comparison of the jamming strain and $Z = D+1$ with the strain values where the system becomes marginal (shown by a black arrow in {\bf (f)}). Different thresholds are consistent with each other. {\bf (c)} Small force distributions ($f_c < \langle f_c \rangle$) for different densities in the steady state. {\bf (d)} Exponent value $\theta$ of small force distributions shown as a function of density and compared with ${\frac{1} { \gamma_g} - 2}$, which needs to be smaller than $\theta$ for stability of the packings. {\bf (e)} Small force distribution as a function of strain, shown for $\phi=0.61$, for different windows $Z$ (or $\gamma$) values. SS in the legend indicates steady state $Z$ values. {\bf (f)} Exponent value $\theta$ of the small force distributions shown as a function of strain and compared with ${\frac{1} { \gamma_g} - 2}$, which needs to be smaller than $\theta$ for stability of the packings. The strain values used to plot small force exponents are the strain value at the lower end of the $Z$ window over which they are averaged.  $\gamma_g$ is the power law exponent of $g(r)$, computed from configurations in a window of $Z$ (or $\gamma$) values (see SI Fig. $S5$.  Supporting data for densities $\phi=0.57$ and $\phi=0.58$ are shown in SI Fig. $S6, S7$).}
\end{figure}

In Fig. \ref{figm4}(a), we show distribution of contact forces (whose magnitudes include normal and tangential components) for different densities in the steady state.  These distributions display peaks at finite forces,  indicative of jamming \cite{ohern-2002}, above the density $\phi=0.55$ which thus marks the lower density limit to jamming. Recently, attention has been focussed on the distribution of forces at the small force limit, observed to obey a power law distribution $P(f) \sim f^{\theta}$, which we consider next. The exponent $\theta$, together with the power law exponent $\gamma_g$ governing the near contact singularity of the pair correlation function $g(r)$ have been related through a stability criterion by Wyart and co-workers \cite{wyart2012,lern2013}. Considering only extended mode instabilities, the inequality expressing the criterion for stability is $\gamma_g \geq  \frac{1}{ 2+ \theta}$, whereas considering local buckling modes, the stability criterion is $\gamma_g \geq  \frac{1-\theta_b}{ 2}$. With a view of studying the extent to which these stability criteria may correspond to the jamming thresholds we observe, we compute the small force distributions, for steady state strains, which are shown in  Fig. \ref{figm4}(c). We observe a regime in small forces which is described by a power law, whose slope decreases as density decreases. The behaviour and significance of the distributions at smaller forces than the power law regime is difficult to analyse with confidence, which we discuss briefly in the SI. In Fig. \ref{figm4}(d), we show the small force exponent values as a function of density. We observe that the inequality  $\gamma_g \geq  \frac{1}{ 2+ \theta}$ is observed at higher densities, and approaches an equality as $\phi = 0.55$ is approached, marking it as the lower density limit to shear jamming.  We monitor the evolution of force distributions as a function of strain (approaching the jamming strain from above) to investigate whether the jamming strain corresponds to marginal stability.  In Fig. \ref{figm4}(e)(f), we show small force distributions (evaluated over bins of strain value for statistics) and the small force exponent as a function of strain, for $\phi=0.61$. 
The strain value above which the stability criterion is met is the jamming strain identified earlier, as we show for three densities  Fig. \ref{figm4}(b). Thus, the marginal stability criterion, analysed for frictionless systems \cite{wyart2012} also describes the shear jamming threshold. This remarkable agreement is not {\it a priori} obvious, and prompts theoretical analysis of the stability criterion for frictional systems under the application of shear deformation.

\begin{figure}[]
\hspace*{-0.8cm}
\includegraphics[scale=0.31,angle=0]{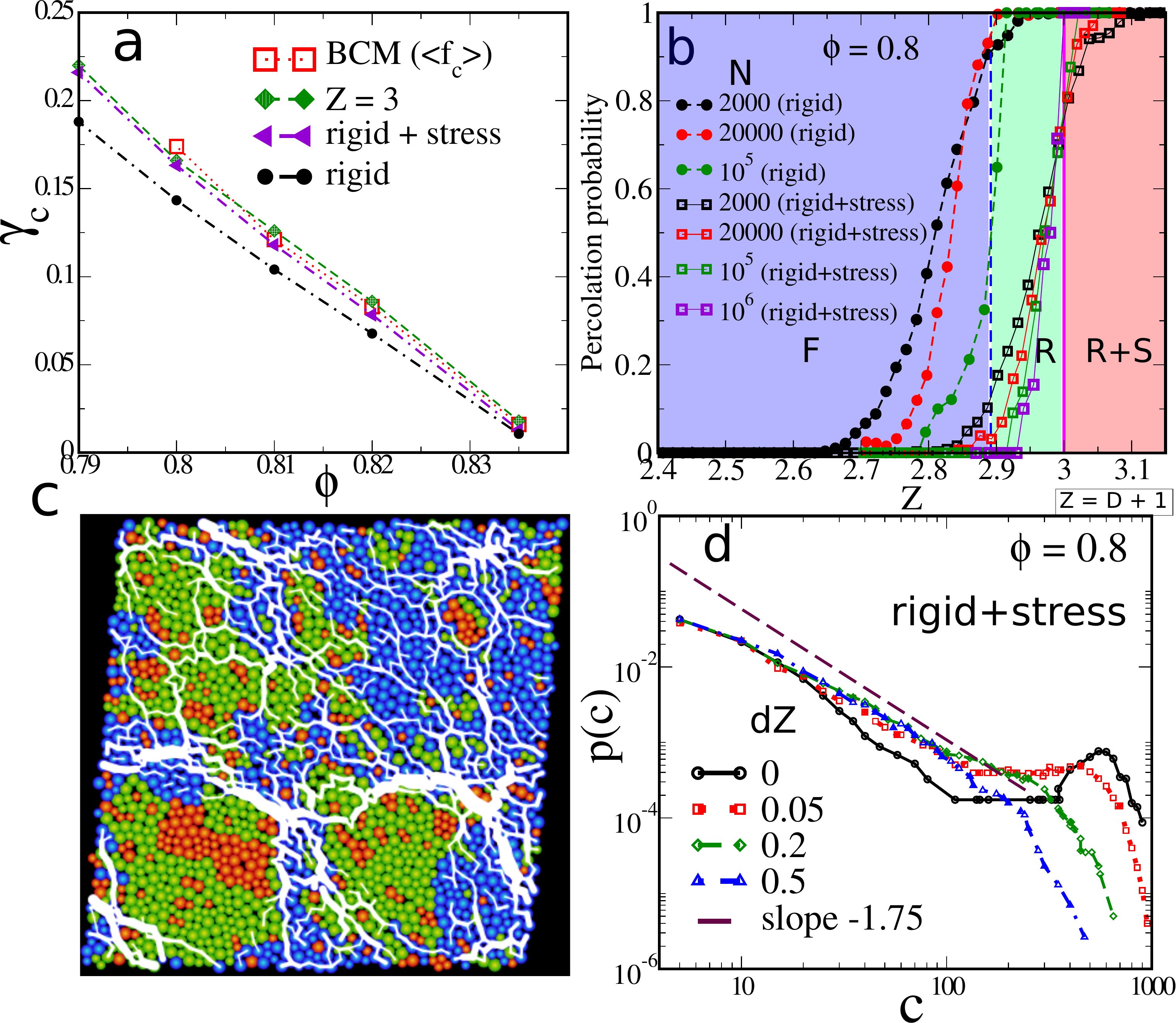}
\caption{\label{figm5} Shear jamming behaviour for the two dimensional soft disc system. {\bf (a)} Threshold shear strains as a function of density.  Data points marked ``$Z = 3$" corresponds to strain values where $Z$ reaches $D+1 = 3$ for each density, and ``BCM" corresponds to the shear jamming strain (see SI Fig. $S9$).  Data points "rigid" and ``rigid + stress"  correspond to strain values where the percolation probability, for percolation along both $x$ and $y$, reaches the value $0.5$ (see SI Fig. $S11$). {\bf (b)} Percolation probability as a function of $Z$, showing the presence of an intermediate phase between rigidity percolation and rigid+stress percolation.  The blue shaded region indicates the floppy (F) phase, the green region indicates the intermediate (R) phase and the red region indicates the stressed or over-constrained phase (R+S).  {\bf (c)} An overlay of the rigid, rigid+stress percolating clusters and the strong force network ($f_c > \langle f_c \rangle$; white bonds), at $\phi=0.82,\gamma = 0.082, Z = 2.928$. Orange discs belong to the floppy regions (or small rigid clusters of size smaller than $5$), green discs belong to the percolating rigid cluster, and blue discs belong to over-constrained regions. The graphic illustrates a strong correlation between the strong force network and the over-constrained region (In SI Fig. $S15$, we show force network and the over-constrained regions for a series of strain values, illustrating this further). {\bf (d)} Distribution of rigid+stress clusters, shown for $\phi = 0.8$, for different values of $dZ = Z_c - Z$ (with which the different curves are labeled). $Z_c$ is computed for each initial configuration corresponding to the percolation of the rigid+stress cluster. The maroon curve represents a slope of $-1.75$, shown for reference. For comparison, for a self-organized rigidity percolation model $\alpha = -1.94$ \cite{briere2007}.}
\end{figure}


The results above characterise the force networks we obtain and the shear jamming limit density of $\phi = 0.55$.  In order to elucidate further the nature of the force networks we generate, we perform a rigidity analysis, as we describe next. Rigidity percolation analysis \cite{thorpe1983,jacobs1995,mouka-1995} has previously been used to study the rigidity of covalent glass networks and jammed packings with and without friction \cite{henkes2016}. We perform this analysis, along with force balance analysis, for a two dimensional soft disc system, to take advantage of reliable methods for rigidity analysis for two dimensional constraint networks. The shear jamming transition for AQS configurations, for $\mu \rightarrow \infty$ (BCM), occurs at strain values very close to $Z = D+1$, which is consistent with the $3D$ results (see also \cite{zhang2010}), as shown in Fig. \ref{figm5}(a). Rigidity analysis is performed using the pebble game  \cite{jacobs1997,henkes2016}, as described in the Methods section. In this analysis, jamming corresponds to the emergence of a system spanning rigid cluster.
In Fig. \ref{figm5}(a), we show that percolation of rigid clusters, along all directions, (marked {\it rigid}) occurs before the isostaticity condition, $Z=(D+1)$, is reached, as has also been observed for sheared frictional packings in \cite{henkes2016}. To understand this discrepancy, we also consider percolation of over-constrained (marked {\it rigid+stress}) regions in our configurations. It has been noted by Moukarzel et al. \cite{mouka-1995}  that the onset of stress transmission through a lattice of springs occurs when stressed clusters of macroscopic size are present. In Fig. \ref{figm5}(a), we observe that strain values of rigid+stress percolation closely correspond to the shear jamming transition. We perform a system size analysis of percolation probabilities, as a function of contact number $Z$, and find threshold values of  $Z \approx 2.89$ for rigid, and $Z \approx 3.0$ for rigid+stress percolation, see Fig. \ref{figm5}(b). Thus, shear jamming corresponds to rigid+stress percolation, preceded by an {\it intermediate} regime of rigidity percolation without stress propagation or shear jamming.  This observation has not previously been reported, although results that suggest such an intermediate regime have been reported for compressed granular packings \cite{bandi}. The presence of an intermediate phase was previously observed in chalcogenides and oxide glasses \cite{intermediate,thorpe2000,chuby2006,briere2007}. In the case of shear jamming, the role and implications of the intermediate phase is not clear and merit further investigation. An appealing possibility is that fragile force networks form in the intermediate phase, see SI Fig. $S13$ and $S14$. To see the relation between force balance conditions and rigidity percolation analysis, we overlay the strong force network $f_c > \langle f_c \rangle $ onto the network of rigid and floppy regions, and observe that contacts with strong forces are mostly concentrated on discs that belong to the over-constrained regions, see Fig. \ref{figm5}(c) and SI Fig. $S15$. We further characterize the nature of the shear jamming transition by the cluster size distribution of rigid+stress percolation, see Fig. \ref{figm5}(d), which show features characteristic of a continuous percolation transition.

In summary, we develop a new approach to solving force balance conditions which improves the accuracy of such calculations, and show that sheared frictionless spheres evolve self-organized structures that can support external stress, and be jammed, should frictional forces also be present. We show convincingly that the random loose packing density of $0.55$ is the low density limit of shear jamming. We find that the mean contact number required for shear jamming is $Z = D + 1$, the isostaticity condition for frictional particles, independent of friction coefficient and shear jamming protocols. This result appears to be valid for shear jamming, but not in general for frictional jamming \cite{makse2008}. We also show that the stability criterion proposed by Wyart  \cite{wyart2012} is valid for the shear jammed states we investigate. This is particularly interesting as the role of near contact neighbours in the analysis of Wyart {\it et al.} \cite{wyart2012,lern2013} also appears to hold for shear jammed states. We compare our analysis of force balance conditions with rigidity percolation analysis and find that rigidity percolation precedes shear jamming in strain at any given density. More interestingly, we show that shear jamming corresponds to the percolation of over-constrained regions, implying also the presence of an {\it intermediate phase} in shear jamming systems, analogously to the case of covalent glass formers. Our results thus reveal many interesting geometric aspects of shear jamming in sharp detail, although the implications of some of these features require further investigation to elucidate. 


{\bf Methods:}

The starting point for our analysis is the specification of the contact network for athermally sheared sphere and disc configurations,  which we generate using shear deformation of frictionless mono-disperse soft spheres (and binary soft discs, specified below) using the athermal quastistatic shearing (AQS) protocol, as detailed in previous work  \cite{vinu2016,vinu-sjperc}, which we refer to for further details. The contact networks obtained from sheared frictionless soft spheres/discs are used to obtain force balance solutions. We study sphere packings for a wide range of densities [$0.55 - 0.63$] and system sizes $N=256, 2000$.  The number of samples used to obtain  Fig. \ref{figm1} and Fig. \ref{figm2} is between $5$ and $10$.   For Fig. \ref{figm3}(a), the data for $N=256$ are averaged over all independent solutions of $5$ steady state (SS) configurations (by which we mean the large strain regime in which the properties of the packings, {\it e. g.} the mean contact number, do not statistically change with changes in strain)   and for $N=2000$ independent solutions are obtained from $6$ configurations. The number of samples used to obtain Fig. \ref{figm3}(b) is above $5$. 
The number of samples used to obtain Fig. \ref{figm4}(a),(c) is $4000$ force configurations and $20$ contact geometries. Force distributions for each $Z$ window in Fig. \ref{figm4}(e) is obtained by averaging at least $20$ contact geometries and more than $50$ force configurations. 
In order to perform rigidity analysis, we analyze sheared frictionless packings in 2D of a $50:50$ binary mixture of soft discs,  the diameter ratio being $1.4$, for densities $0.79 - 0.835$ and system sizes ($N=2000,20000,10^{5},10^{6}$). The number of configurations used to obtain percolation data in Fig. \ref{figm5} are $300,75,10,10$ for $N=2000,20000,10^{5},10^{6}$ respectively.


{\bf Force Balance Solutions -- The null space method:} We develop a new method based on projecting the problem onto the null space of the contact matrix, described below. Various simulation techniques and numerical methods have been employed in previous work to obtain force balanced contact geometries, and force networks for a given contact geometry \cite{lerner2012,lerner2013,gendelman2016,jean1999,unger2003,hurley2016,snoeijer2004}. Our method vastly improves the accuracy of the solutions, see SI Fig. $S1$(b). To obtain contact forces for a given contact network, first we express the vector $\vec{r}_{ij}$ joining the center of two spheres $i$ and $j$ in spherical polar coordinates ($\hat{n}_{ij},\hat{\theta}_{ij}$,$\hat{\phi}_{ij}$). Let $\vec{f_i}$ and $\vec{\Gamma_i}$ denote the total force and the total torque on the $i^{th}$ particle and $\vec{f_{ij}}$ is the force exerted on particle $i$ from particle $j$. Then the force and torque balance conditions are written as follows.
\begin{eqnarray}
 \vec{f_i} = \Sigma _j (\hat{n}_{ij} f_{ij}^{n} + \hat{\theta}_{ij} f_{ij}^{\theta} + \hat{\phi}_{ij} f_{ij}^{\phi}) = 0 \\
 \vec{\Gamma_i} = \Sigma_j \vec{R_i} \times \vec{f_{ij}} = 0 \\
 \vec{\Gamma_i} = \Sigma_j R_i ( f_{ij}^{\theta} \hat{\phi}_{ij} - f_{ij}^{\phi} \hat{\theta}_{ij}) 
\end{eqnarray}
Where $R_i$ is the radius of particle $i$. The matrix $M$ is constructed from the unit vectors between particles in contact. For a single contact ( SI Fig. $S1$(a)) the matrix $M$ is shown below. 

\begin{equation}
M= \left(
\begin{array}{cccc}
    n_{12}{}^x & \theta_{12}{}^x & \phi_{12}{}^x  & \cdots \\
    -n_{12}{}^x & -\theta_{12}{}^x & -\phi_{12}{}^x & \cdots  \\

      n_{12}{}^y & \theta_{12}{}^y & \phi_{12}{}^y  & \cdots \\
    -n_{12}{}^y &  -\theta_{12}{}^y &  -\phi_{12}{}^y & \cdots \\

     n_{12}{}^z & \theta_{12}{}^z & \phi_{12}{}^z  & \cdots \\
    -n_{12}{}^z & -\theta_{12}{}^z & -\phi_{12}{}^z & \cdots \\

     \vdots  & \vdots & \vdots & \vdots \\ 
     
     0  & R_1 \phi_{12}{}^x & -R_1 \theta_{12}{}^x & \cdots \\
    0 & R_2 \phi_{12}{}^x & -R_2 \theta_{12}{}^x & \cdots \\

   0 & R_1 \phi_{12}{}^y & -R_1 \theta_{12}{}^y & \cdots \\
   0 & R_2 \phi_{12}{}^y & -R_2 \theta_{12}{}^y & \cdots \\

   0 &  R_1 \phi_{12}{}^z & -R_1 \theta_{12}{}^z & \cdots\\
  0 &   R_2 \phi_{12}{}^z & -R_2 \theta_{12}{}^z & \cdots  \\
  
  \vdots  & \vdots & \vdots & \vdots \\ 
\end{array}
\right)
\label{eq:M}
\end{equation}
Then, we write down the force balance and torque balance conditions in the matrix form.
\begin{equation}
  M \mid F \rangle = 0
\end{equation}
 where $M$ is $((\frac{D(D+1)}{2})N \times DC)$ matrix, $D$ is the dimensionality of space, $C$ is the number of contacts and $\mid F\rangle$ is a vector of size $DC \times 1$, with $3$ (for $D=3$, 2 for $D=2$) force components ($f^n, f^{\theta}, f^{\phi}$) for each contact. The normal forces $f^n$, which form the first $C$ elements of the matrix, need to be positive, which is a constraint to be imposed on all solutions.  Now, we construct a matrix $H = M^{T} M$, which is of dimension $DC \times DC$. Using the matrix $M$, we construct an energy function $E = \langle F\mid M^{T}M \mid F \rangle$ to directly obtain contact forces ({\it Direct Minimization} method). But the null space method we describe here offers a more accurate method. To implement this method, we diagonalize the matrix $H$ and obtain eigenvectors with zero eigenvalues, as they satisfy conditions of mechanical equilibrium, which form the basis of null space. Hence any vector (force solution) obtained from a linear combination of these eigenvectors in the null space also satisfies the force balance conditions. Let ${\bf X_j}$ represent eigenvector $j$  and has $DC$ elements. The  first $C$ elements do not in general satisfy positivity constraints and hence are not physical forces. We obtain physical force solutions in the null space by finding the coefficients $x_j$ of the eigenvectors that will satisfy the positivity constraint on $f_n$.  In other words: 
\begin{equation}
 f_{i} :   \Sigma_{j=1}^{D_{ns}} X_{ij} x_{j} - y_i  = 0,  y_i  \geq  0  
\end{equation} 
where $i$ represents the contact number, $D_{ns}$ is the number of eigenvectors in the null space, and $y_i$ are auxiliary variables introduced to impose positivity.  The above set of equations can be written in a matrix form.
\begin{equation}
M^{'} \mid  x y \rangle  = 0, 
\end{equation}
where the matrix $M^{'}$ has the dimension $C \times (D_{ns} + C)$, the vector $\mid xy \rangle$ has dimension $(D_{ns}+C) \times 1$ and are given by
\begin{equation}
M^{'}= \left(
\begin{array}{ccccccc}
    X_{11} & X_{12} & \cdots  & X_{1k} & -1  &  0  & \cdots  \\
    X_{21} & X_{22} & \cdots  & X_{2k} & 0   & -1  & \cdots  \\  
    \vdots  & \vdots & \vdots & \vdots & \vdots & \vdots & \vdots \\ 
    X_{C1} & X_{C2} & \cdots  & X_{Ck} & \cdots &  0 & -1   \\
       
\end{array}
\right)
\label{eq:M'}
\end{equation}
and
\begin{equation}
 \mid xy \rangle = \left(
 \begin{array}{c}
  x_{1} \\ : \\
  : \\ x_{k} \\ 
  y_{1} \\  : \\ 
  : \\ y_{C}\\
 \end{array}
\right )
\end{equation}
By defining an energy function $E^{'} = \langle xy \mid M^{'T} M^{'} \mid xy \rangle$ which is a quadratic function with positivity constraints on $y_i$. The energy functions $E^{'}$ and $E$ are minimized using reflective Newton method for bound constraint minimization (BCM) \cite{coleman1996}, and we use the label {BCM} to refer to either one of the methods above.  We find that independent initial guesses generate independent (but not {\it orthogonal}) solutions. We use random forces as well as forces from DEM simulations as initial guesses. The force scale is set by the magnitude of the initial guess. For SS packings, using the null space method, we obtain all the independent solutions for different densities.\\

We also implement the Coulomb criterion that restricts the magnitude of the tangential forces, in two and three dimensions. We show data for $\mu=1$, a physical value of the friction coefficient,  in addition to solutions for the infinite friction case.\\
In $2D$, the Coulomb criterion, $\mid f_t \mid  \leq \mu f_n$, can be written as a set of linear constraints as follows:
\begin{eqnarray}
f_t - \mu f_n \leq 0 \\
- f_t - \mu f_n \leq 0 
\end{eqnarray}
We include these constraints in the matrix $M$ or $M^{'}$, by introducing auxiliary variables as above. 

 In 3D, the Coulomb criterion is a quadratic constraint, but can be expressed as linear constraints as follows. We have 
\begin{eqnarray}
\lVert  f_t \rVert \leq \mu f_n \\
(f_{\theta}^2 + f_{\phi}^2 ) \leq \mu^2 f_n^2 \\
f_{\phi}^2 \leq (\mu f_n  - f_{\theta}) (\mu f_n + f_{\theta}) 
\end{eqnarray}
which can be written as 

\begin{eqnarray}
\mid f_{\theta}\mid \leq \mu f_n \\
\mid f_{\phi} \mid  \leq (\mu f_n - f_{\theta})\\
\mid f_{\phi}\mid  \leq (\mu f_n  +  f_{\theta}) 
\end{eqnarray}
The last three equations are implemented in a similar way as the $2D$ case. This increases the dimension of the matrix $M$ or $M^{'}$ by $2\times (3C)$ and hence it is computationally expensive.  The quadratic constraint can be directly imposed and solved using interior point methods developed for solving quadratically constrained quadratic program (QCQP). Solutions for these constrained optimization problems are obtained using the optimization library in MATLAB.\\

Using the above method, we first obtain force balance solutions using the forces obtained from DEM simulations \cite{cundall1979} for the same initial SS configurations. In SI Fig. $S2$, we show that force solutions obtained from the two methods match well.\\\\

{\bf Stress anisotropy:} Stress anisotropy is computed from the eigenvalues of the stress tensor. Stress tensor is defined as follows:
\begin{equation}
\hat{\sigma} = \frac{1}{V}\Sigma_{i \neq j} \vec{r}_{ij} \otimes \vec{f}_{ij}
\end{equation}
Where $\vec{r}_{ij}$ is the distance between the centers of spheres $i,j$.
We diagonalize the stress tensor and obtain principal components of the stress tensor and the corresponding eigenvalues. Eigenvalues of the stress tensor are $P_1 > P_2 > P_3$, where $P_1$ is along the compressive direction, $P_2$ is along the transverse or vorticity direction, and $P_3$ is along the dilative direction. The stress anisotropy is defined as follows:

\begin{equation}
 SA = (P_1 - P_3)/(P_1 + P_2 + P_3)
\end{equation}

{\bf The pebble game algorithm:} We implement a ($k=3,l=3$) pebble game \cite{jacobs1997,henkes2016}, which we describe briefly. The algorithm is based on Laman's theorem, which states that the network of $N$ vertices is generically, minimally rigid in two dimensions if and only if it has $3N - 3$ bonds and no subgraph of $n$ vertices has more than $3n - 3$ bonds \cite{laman1970}. The pebble game is implemented as follows: Each disc (from now on called as a site) is assigned $k=3$ pebbles, each pebble corresponding to one degree of freedom,  $2$ translational and $1$ rotational. The quantity $l=3$ corresponds to the total number of global translational and rotational degrees of freedom that are always present for a rigid body and must be accounted. Each contact represents one constraint for translational motion and one constraint for rotational motion. Hence, each contact in the contact network is replaced by two bonds, representing the two constraints. A series of steps are used to assign pebbles to bonds, such that each bond that is covered by a pebble is an independent constraint restricting one degree of freedom. When a pebble is assigned to a bond, it is assigned a direction,  If the bond $E_{ab}$ is covered by a pebble from site $a$, then the bond is directed from $a$ to $b$. Pebbles that remain on the sites are free and can be used to cover other bonds, but once a bond is covered by a pebble, it continues to be covered, and a pebble covering a bond can be moved only with another pebble taking its place.  A bond can be covered by a free pebble from one of the two sites at either end of it, if the total number of pebbles at its sites is $(l+1)$.  If for a bond the total number of pebbles at its sites is less than $l+1$, we search for a free pebble to cover this bond, but the search is conducted only along bonds that are directed away from the sites adjacent to the bond ({\i. e.}, we attempt to retract a pebble previously assigned to a bond, and replace it with another from elsewhere).  From each site $v$, we can search along ($k=3$) bonds directed away from $v$.   The search for a free pebble continues until a pebble is found and a sequence of swaps allows the bond under consideration to be covered and  marked as an independent bond.  If a free pebble is not found, due to the search process encountering a set of closed loops that takes the search back to the initial sites, then the bond is marked as redundant.   The algorithm terminates when each bond is marked as an independent bond or a redundant bond. The presence of redundant bonds leads to over-constrained regions. The sites visited during the failed pebble search belongs to the over-constrained (rigid +stress) regions.

We map out rigid clusters from the network of redundant and independent bonds. Only the independent bonds are used to identify rigid clusters, and we seek to label them so that all the independent bonds belonging to a rigid cluster has the same cluster label. We start with an unlabelled bond $E_{ab}$ and we perform a pebble search to obtain  $k=3$ pebbles (which can always be found) and pin the three pebbles at the sites $a$ and $b$ (i.e., these three pebbles are not free).  Sites $a$ and $b$ are marked rigid and bond $E_{ab}$ is assigned a cluster label. Next, we inspect neighbours of sites $a$ and $b$ to mark them floppy or rigid with respect to $a$ and $b$. For $a_1$, a neighbour of site $a$, we perform a pebble search and attempt to free a pebble for the bond $E_{a,a_1}$.  If a free pebble is found then site $a_1$ is marked floppy. If a free pebble is not found then the site $a_1$ is marked rigid and also all the sites that are visited during the failed search are marked as rigid with respect to the initial sites. The above procedure is repeated until all the neighbours of the rigid sites are marked floppy. All the bonds between pairs of sites marked rigid are given the same cluster label as $E_{ab}$. The procedure is repeated until all the bonds are assigned a cluster label,  with the labelling as rigid or floppy being removed when a new cluster search is initiated each time. Clusters containing only two sites (one bond) belong to floppy regions. More details of the algorithm are in the reference \cite{jacobs1997}.

After the pebble game, we have sites (or discs) belonging to the biggest rigid cluster and also sites that are identified as over-constrained, which we then test for percolation. The percolation probability is computed by first identifying the biggest cluster by connecting particles that are in contact using Lees-Edwards periodic boundary conditions.  Then we check if the biggest cluster percolates not only in the simulation box (by inquiring whether it extends from one edge of the box to the other, as done in previous studied \cite{vinu-sjperc}),  but also considering an extended system composed of the simulation box, surrounded by periodic copies of the system, and testing whether the largest cluster percolates across the extended area so defined. These two procedures do not produce significantly different results in cases studied before \cite{vinu2016,vinu-sjperc}, but we find that for rigidity percolation analysis, the thresholds depend more significantly on the definition used. We thus use the more restrictive definition of percolation across the extended area. Since the force network in the sheared systems we study are anisotropic, we study percolation independently along  one direction ($x$ or $y$ or compressive (diagonal)) and all directions ($x$ and $y$ or compressive and dilative directions). 

To obtain cluster size distributions, shown in Fig. \ref{figm5}, we first compute the percolation threshold value of the contact number $Z_c$ seperatly for each configuration as the value where where the rigid+stress particles percolate {\it i.e.,} cluster spans at least in one direction. The average value of $ \langle Z_c \rangle = 2.858$, is used in the plot shown.   

\bibliography{sjforbal}
\bibliographystyle{naturemag}

\clearpage

\part*{}   

\setcounter{figure}{0}
\renewcommand{\thefigure}{S\arabic{figure}}%

\begin{center}
\textbf{Force networks and jamming in shear deformed sphere packings(Supplementary Information)}
\\
{H. A. Vinutha and Srikanth Sastry}
\date{\today}
\end{center}

Here we present additional information regarding various aspects of our analysis of force balance solutions, namely: (i) Accuracy of the null space method, (ii) Comparison with DEM forces, (iii) Dimensionality of the null space, (iv) Small force exponents and stability criteria, (v) Small force distributions as a function of strain, (vi) Spatial correlations of stress, (vii) Density-Strain phase diagram for the two dimensional system, (viii) Rigid and rigid+stress percolation in two dimensions, (ix) Intermediate phase in shear jamming and (x) Evolution of rigid, rigid+stress, and force networks. 

\section{Accuracy of the Null Space Method}
\begin{figure*}[] 
\includegraphics[scale=0.36,angle=0]{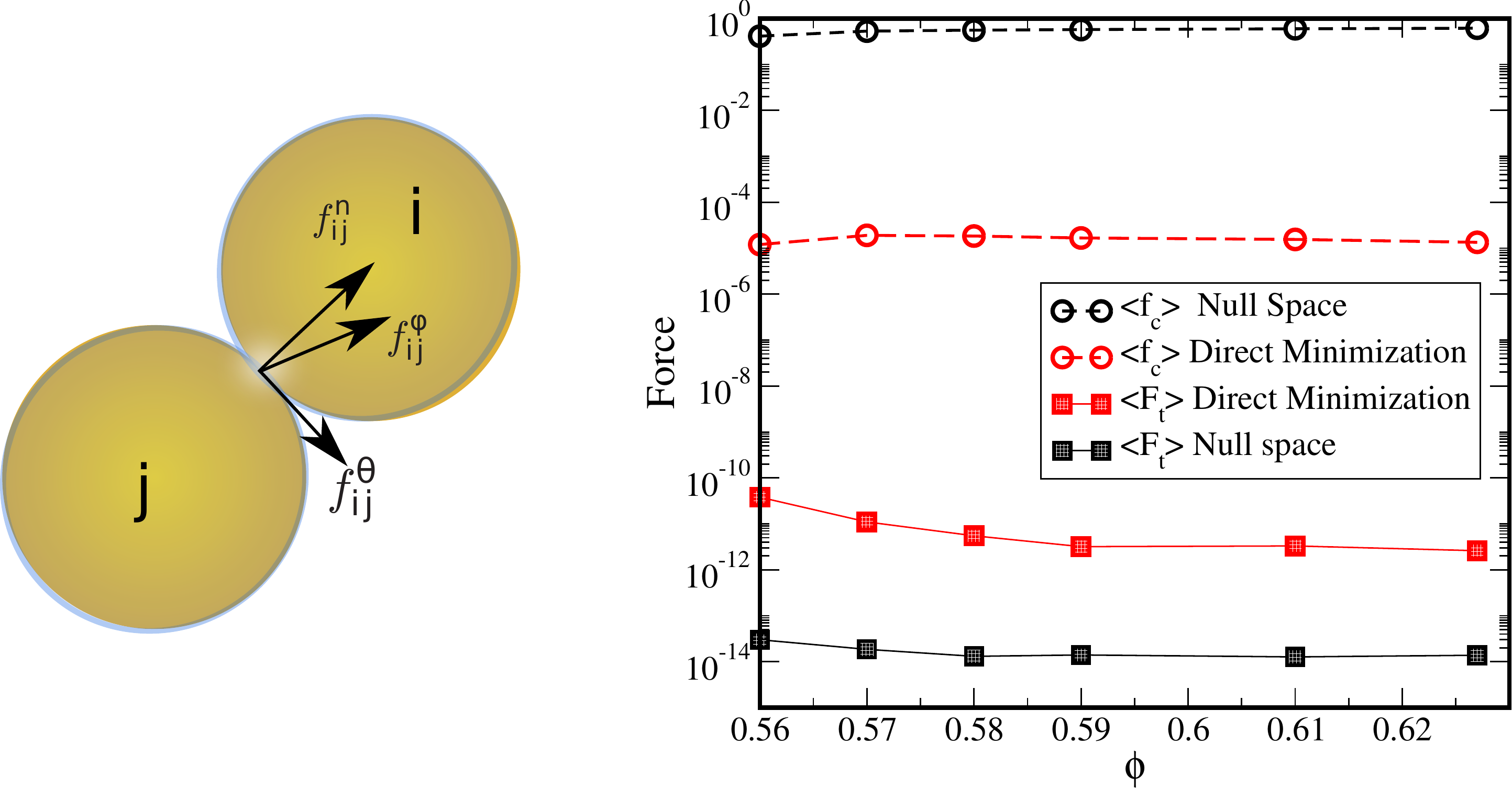} \vspace{-3mm}
\caption{\label{figmeths} {\bf (a)} Illustration of two particles in contact and the forces exerted by particle $j$ on particle $i$. {\bf(b)} Average contact and total forces obtained using the null space method and direct minimization for different densities. The plot shows that the force solutions obtained using the null space method are better force balanced.}
\end{figure*}

In Fig. \ref{figmeths}, we show the average contact force $\langle f_c \rangle$ and the average of total force on a particle $ \langle F_t \rangle$,  for SS configurations at different densities, for force solutions that are obtained using the null space method and from  direct minimization. The magnitude of the average contact force is set by the initial guess used by the minimization protocol. Figure \ref{figmeths} shows that force solutions obtained using the null space method are more accurate and are force balanced up to an accuracy of $\approx 10^{-14}$, whereas the force solutions obtained from direct minimization are force balanced up to an accuracy of $\approx 10^{-11}$.

\section{Comparison with DEM forces}
Using the BCM method, we first obtain force balance solutions and compare the solutions to the forces obtained from DEM simulations for the same initial SS configurations, obtained quasistatically with a small strain step to ensure there is a minimal change in the contact network during DEM \cite{vinu2016}.The initial guess for forces is taken from DEM simulations, and the simulation details and DEM parameters are as mentioned in \cite{vinu-sjperc}. The normal and tangential components of the contact forces, $f_n$ and $f_t = \sqrt((f^{\theta})^2 + (f^{\phi})^2)$, are computed for densities $\phi=0.627,0.61$, and $0.58$, including in the matrix $M$ only those contacts that remain at the end of the DEM simulations. The forces at each contact estimated using BCM are compared with those obtained from DEM simulations in Fig. \ref{figp1}, demonstrating that they agree very well with each other. This analysis supports two conclusions: (i) Forces obtained by solving force balance conditions for SS configurations agree well with DEM forces, implying that the SS contact network can support the forces observed in the DEM simulations, which in turn correspond to shear jamming. (ii) The contact networks at the end of the DEM simulations, which support finite contact forces and stresses, are those that are present in the SS configurations. The only changes in the contact network is that some of the SS contacts are lost during the DEM simulations, and we can account for all of them. They are either contacts of rattlers, or contacts that require higher friction coefficients to be retained than what we employ during the DEM simulations.

\section{Dimensionality of the Null Space}
\begin{figure}[h!]
\includegraphics[scale=0.4,angle=0]{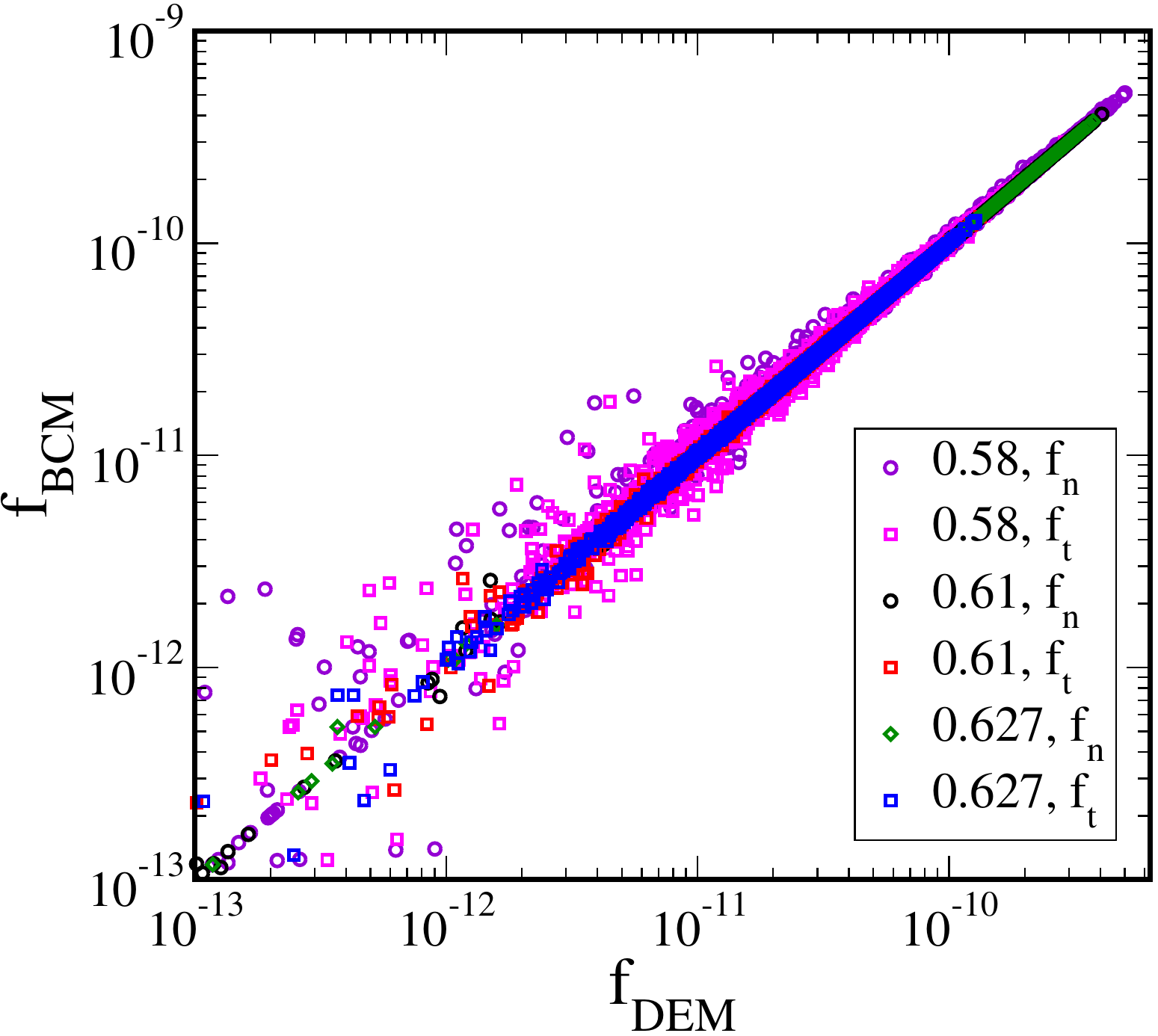} \vspace{-4mm}
\caption{\label{figp1} Comparison of forces from DEM simulations and BCM, shown for different densities. $f_n$ and $f_t$ are the normal and the tangential contact forces.} \vspace{2mm}
\includegraphics[scale=0.45,angle=0]{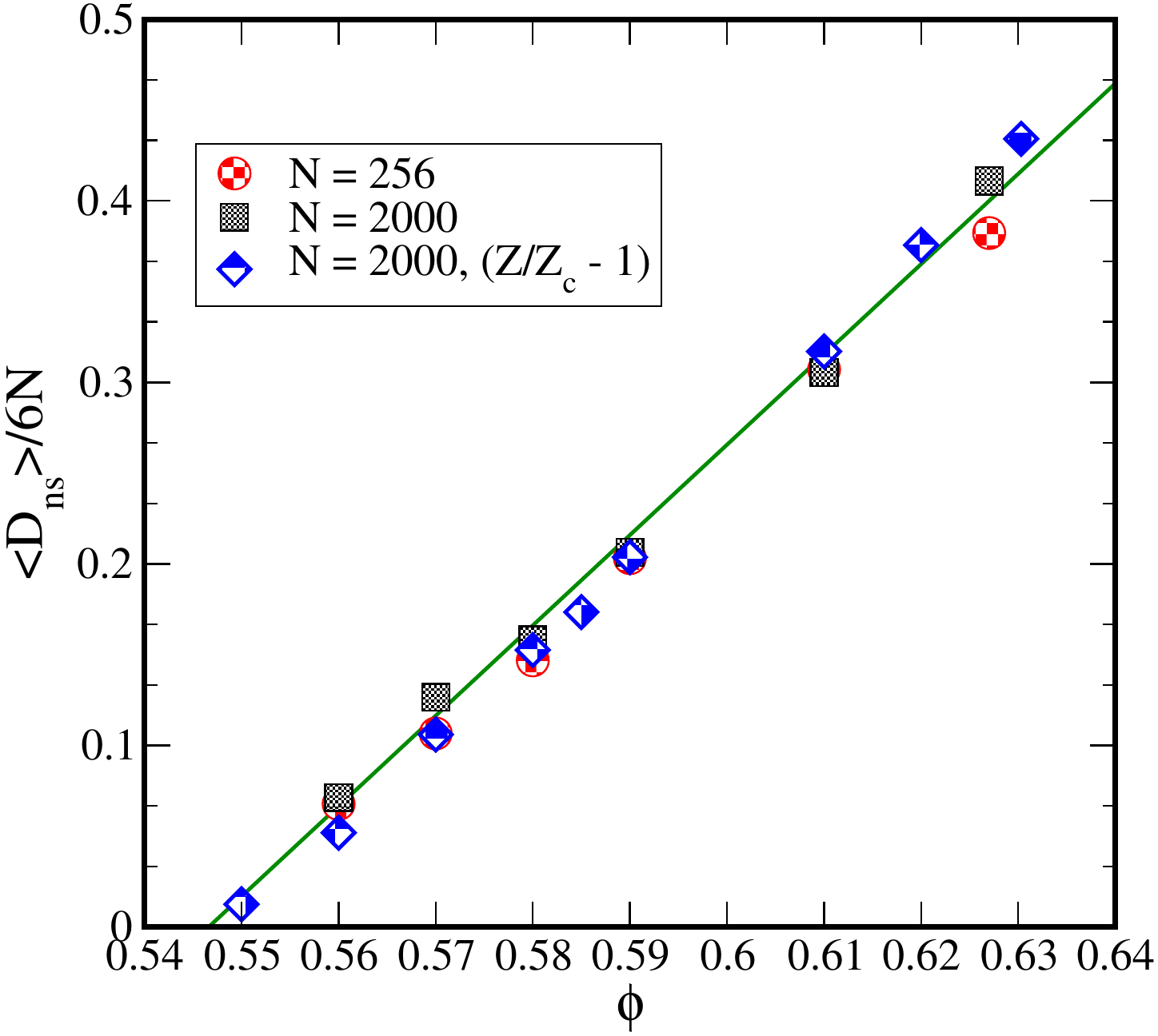} \vspace{-3mm}
\caption{\label{fig2} The null space dimension $D_{ns}$, scaled with $6N$ (maximum dimension of the matrix $M$), as a function of density for two different system sizes. The green line is a fit to $N=2000$ data (black squares). The plot shows that the fit curve and the $Z$ data of SS configurations identifies $\phi=0.55$ as the lower density limit, as $Z$ approaches the isostatic limit at $\phi=0.55$, $Z_c = 4 (= D+1)$.}
\end{figure}

In Fig. \ref{fig2}, we show the null space dimension as a function of density for $N=256$ and $2000$, averaged over $5$ initial SS configurations. Observe that the limiting density when the number of force solution is one is close to $\phi=0.55$, indicated by the green fit curve for the $D_ns/6N$ ($N=2000$) data. This is consistent with other indicators discussed in the paper.

\section{Small Force Exponents and Stability Criteria}
\begin{figure*}[h!] 

\includegraphics[scale=0.45,angle=0]{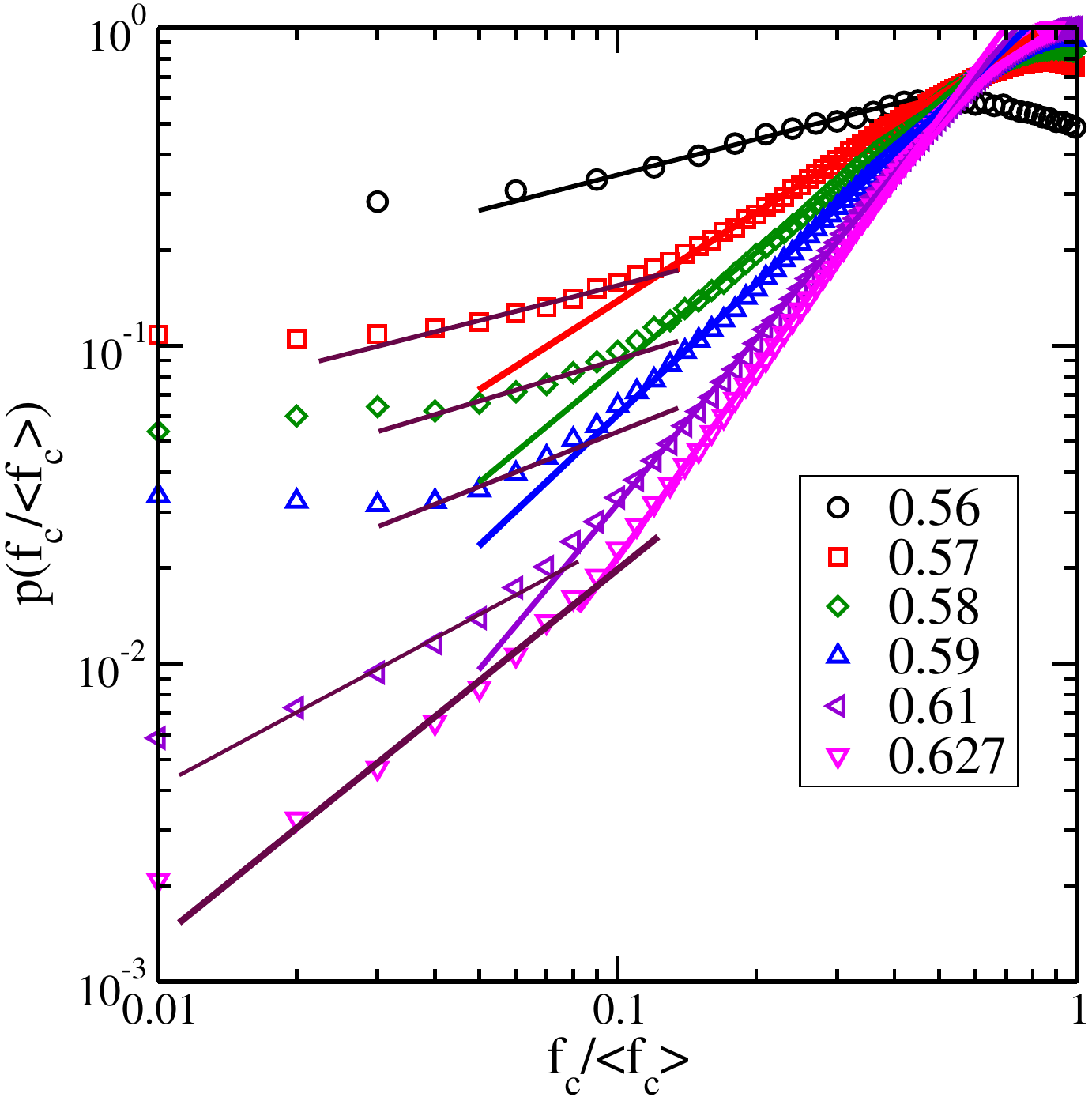}
\hspace{0.5cm}
\includegraphics[scale=0.45,angle=0]{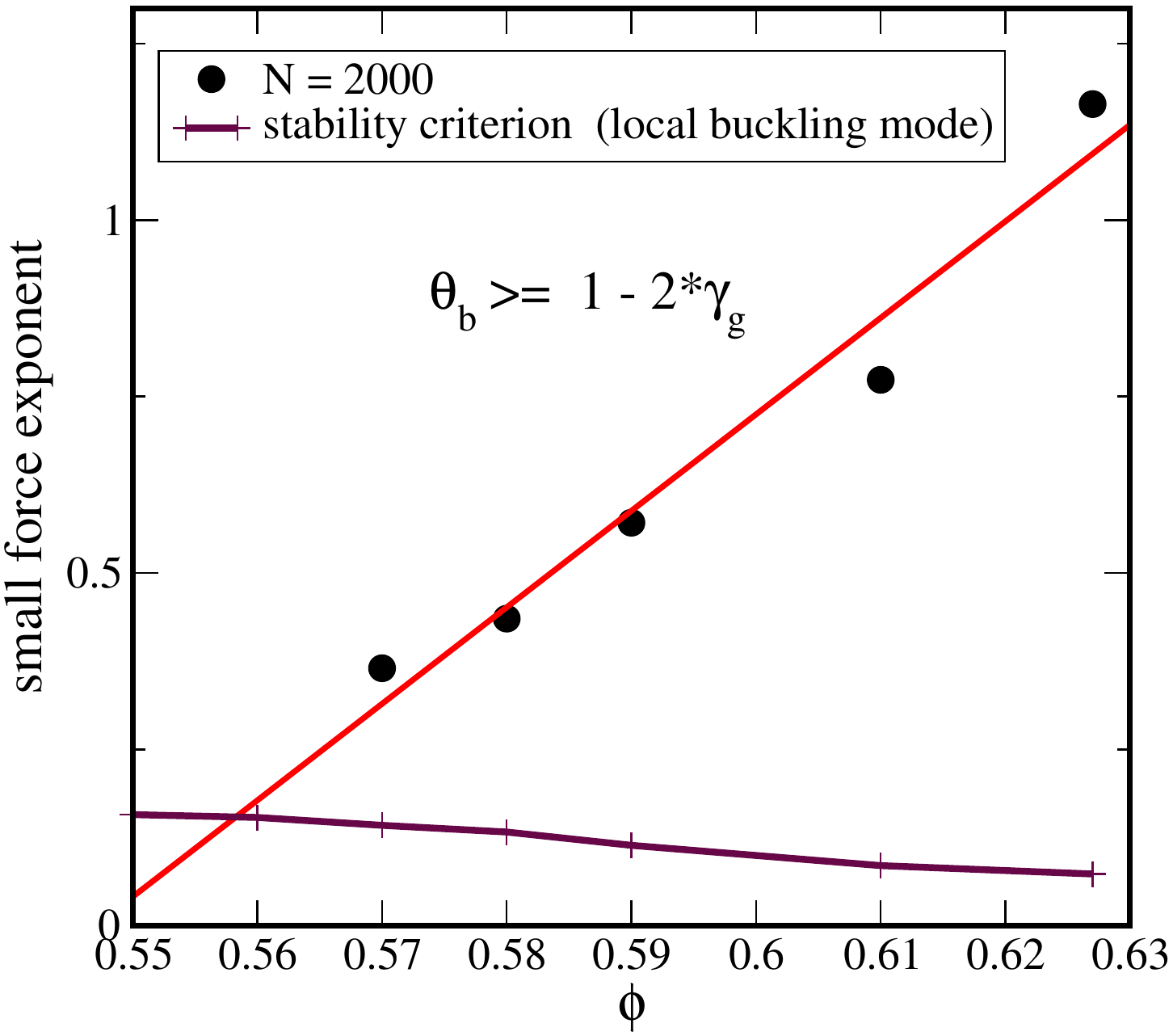}
\caption{\label{figbm} {\bf(a)} Small force distributions as a function of density. {\bf(b)} Small force exponent $\theta_b$ as a function of density .}
\end{figure*}

In Fig. $1$(d) of the paper, we showed force distributions obeying a power law distribution $P(f) \sim f^{\theta}$. The exponent $\theta$ and the power law exponent $\gamma_g$ of the near contact singularity of the pair correlation function $g(r)$ have been related through the stability criterion $\theta > \frac{1}{\gamma_g} - 2 $. In Fig. $3$(b) of the paper only the variation of small force exponent $\theta$ is shown, which corresponds to extended mode instabilities. In addition to extended modes, a small force distribution arising from buckling modes has also been discussed. We find in our force distributions that the smallest force regime is characterised by a smaller exponent, which we tentatively identify as the buckling mode exponent  $\theta_b$, which is obtained by a fit to small contact forces (the fit curves  in maroon, see Fig. \ref{figbm} (a)). In Fig. \ref{figbm}, we show the variation of the small force exponent $\theta_b$, as the lower density limit is approached. Similar to the $\theta$ exponent, $\theta_b$ also shows that as the lower density limit of $\phi = 0.55$  is reached the system ceases to obey the stability criterion. In table $1$, we tabulate different exponent values corresponding to extended and buckling modes and the exponent values obtained from relations of marginal stability. Note that we do not have data to directly substantiate the association of forces as due to extended or buckling modes.

\section{Small force distributions as a function of strain}
\begin{figure*}[htbp!] 
\includegraphics[scale=0.45,angle=0]{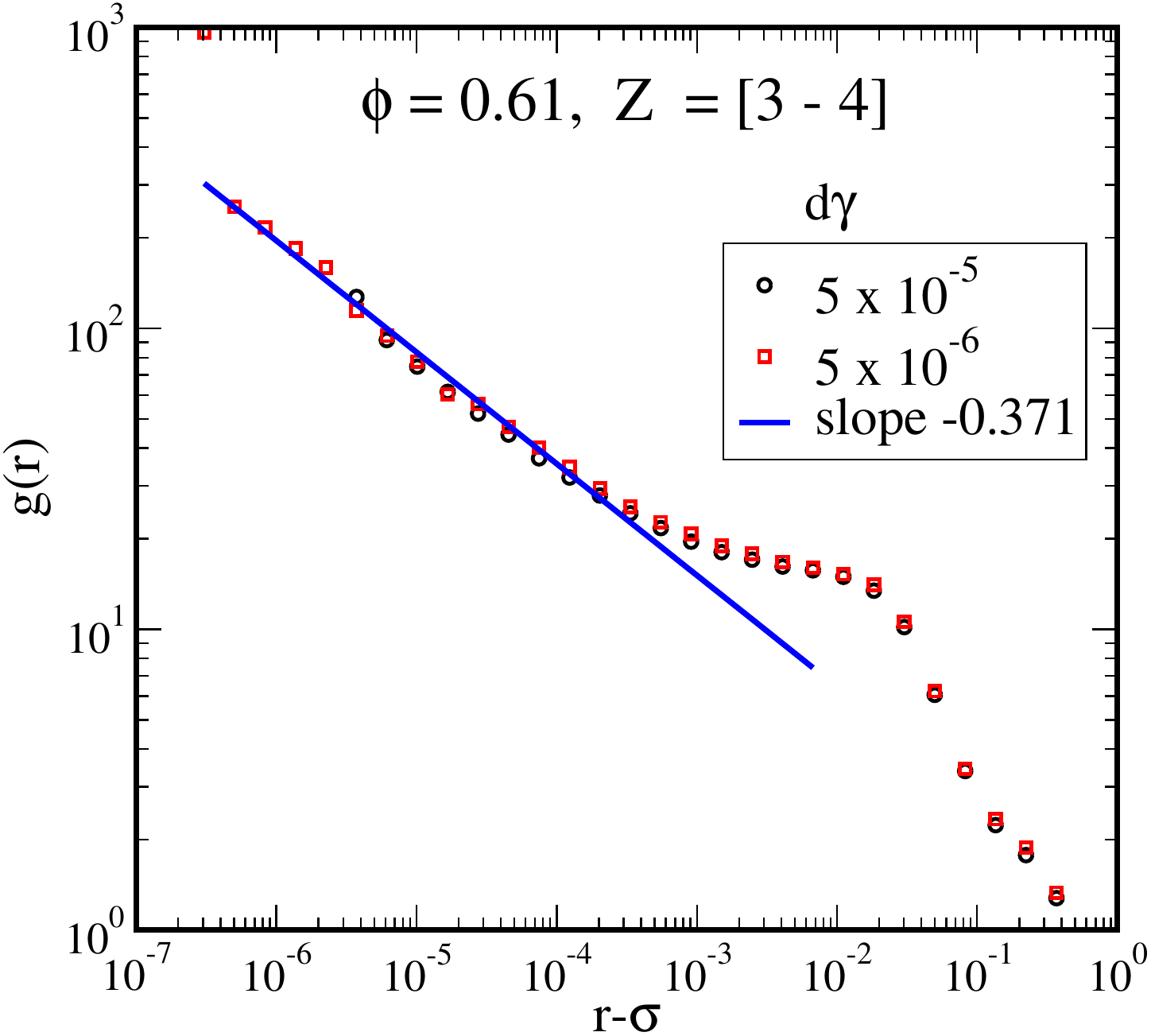}
\includegraphics[scale=0.45,angle=0]{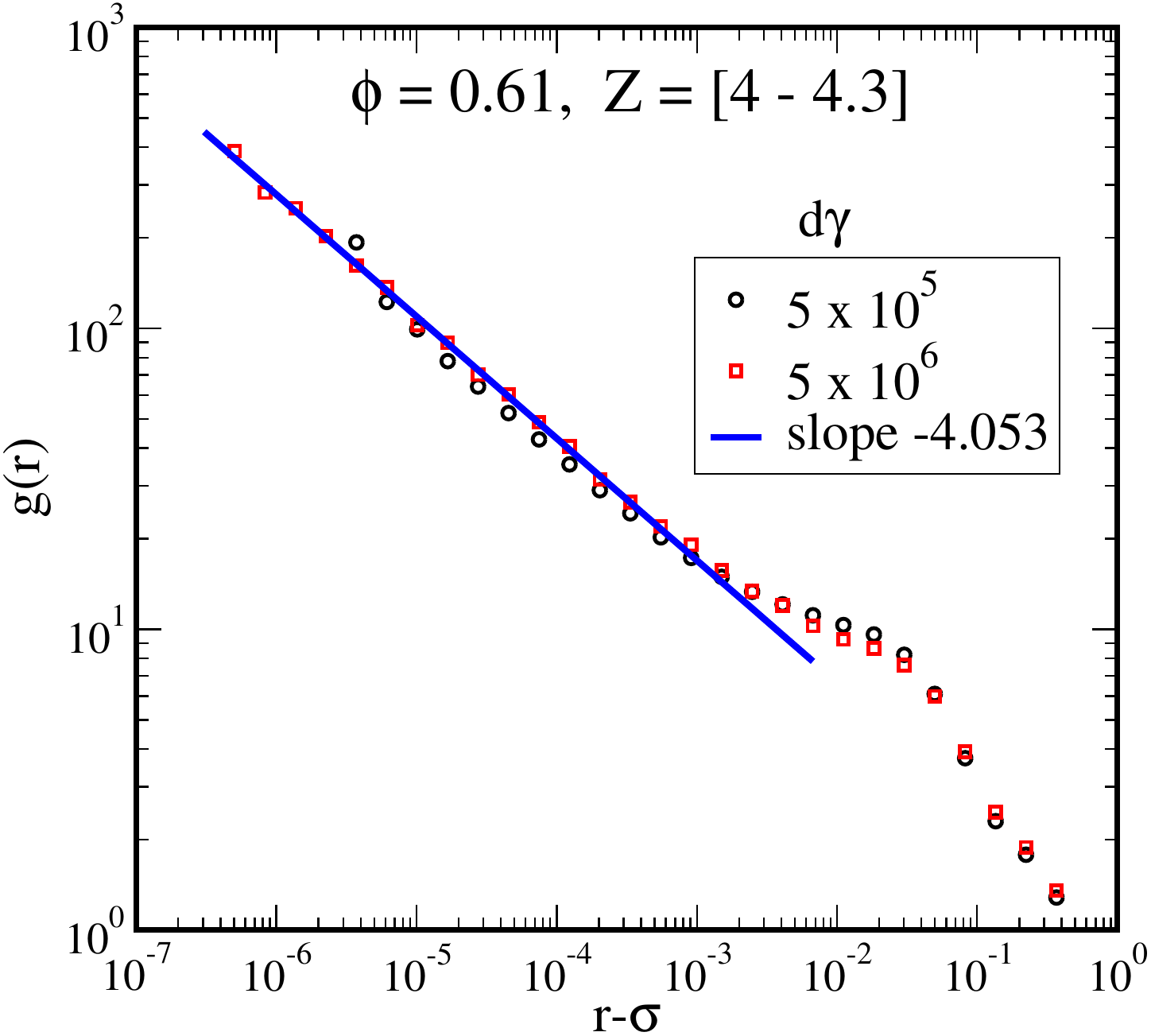}
\includegraphics[scale=0.45,angle=0]{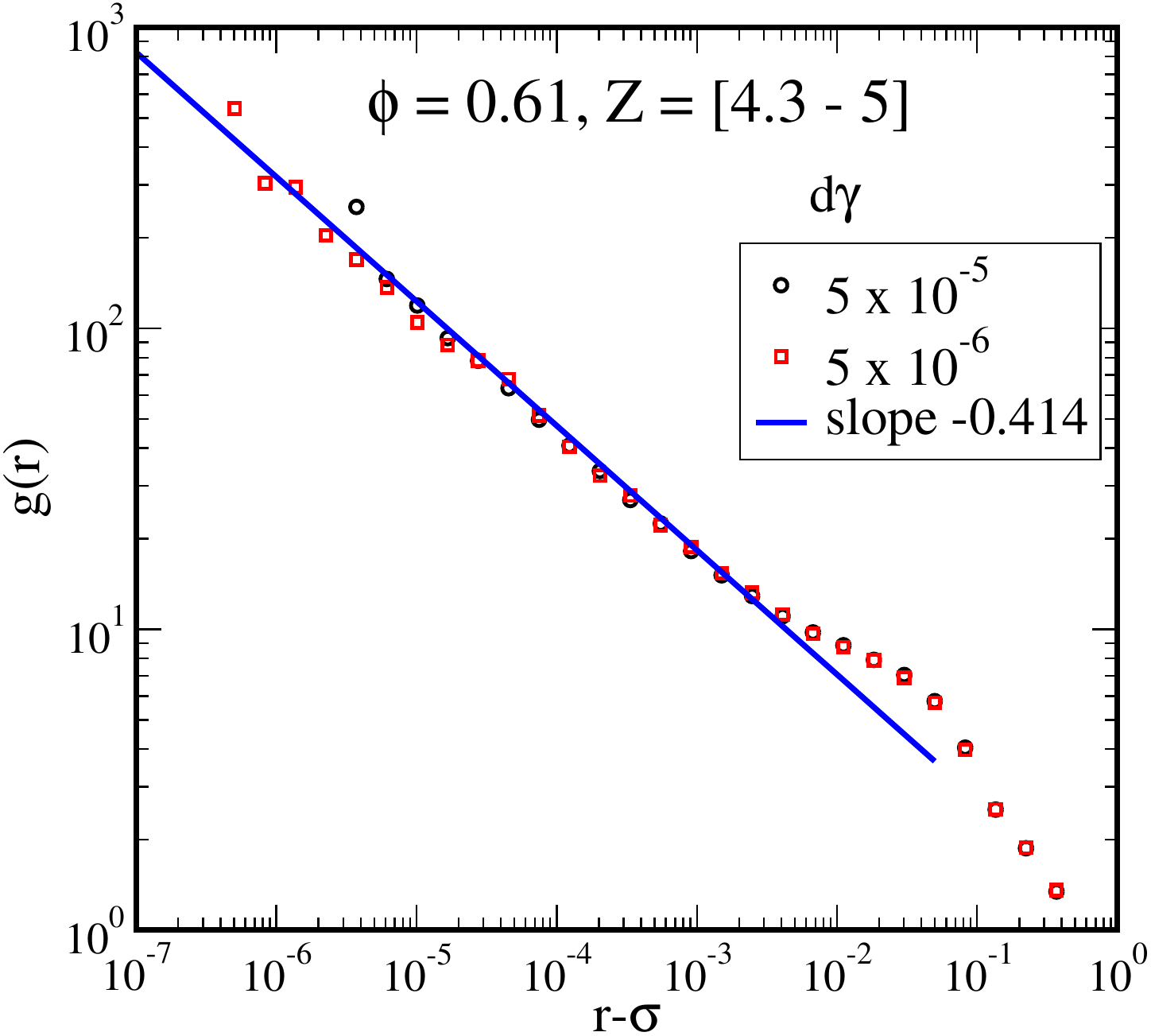}
\includegraphics[scale=0.45,angle=0]{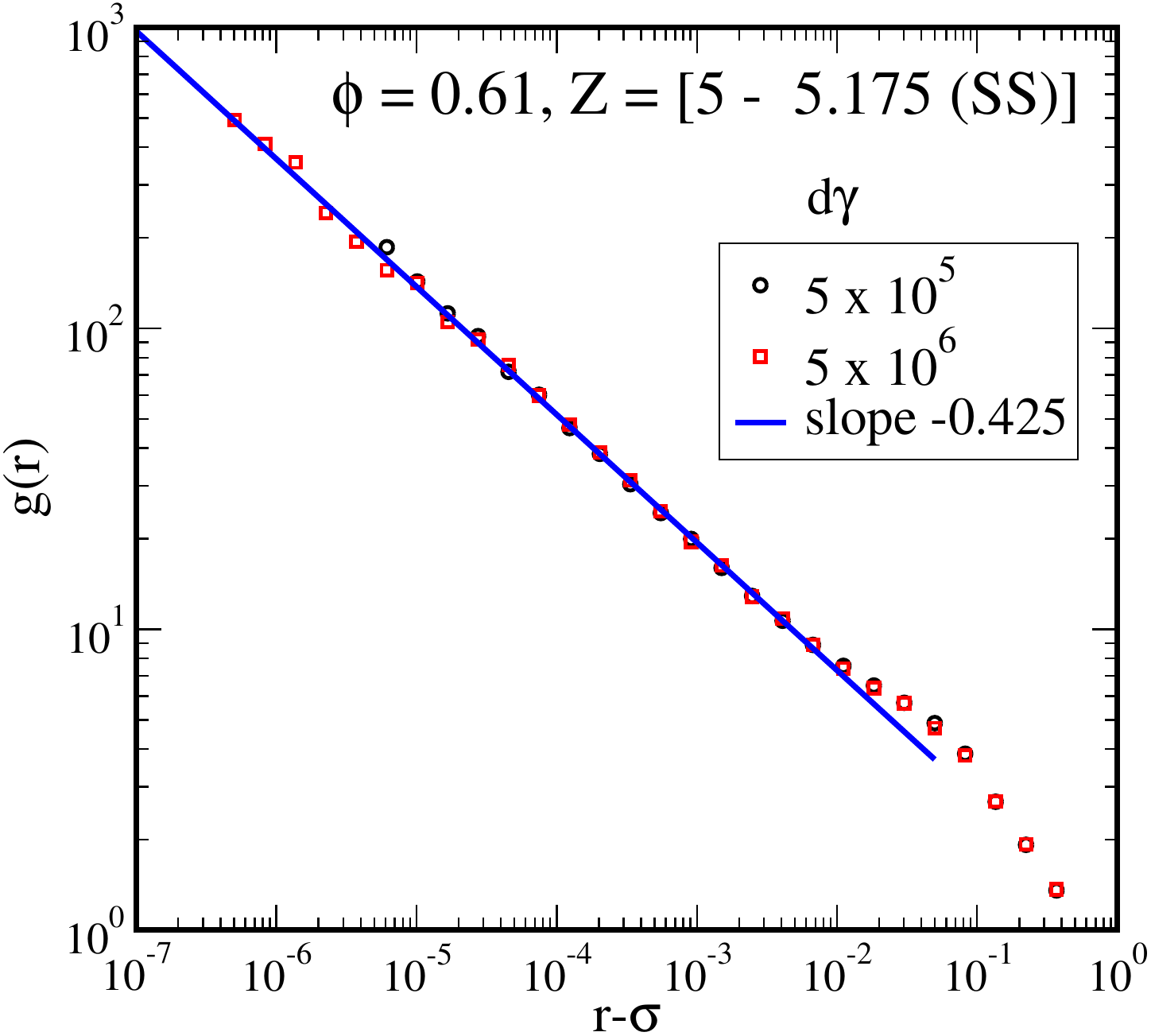}
\caption{\label{figgofr} Near contact power law form of $g(r)$ for different windows of $Z$ (or $\gamma$) values, {\bf(a)} $Z = [3 - 4]$ and $\gamma = [0.055 - 0.088]$, {\bf(b)} $Z = [4 - 4.3]$ and $\gamma = [0.088 - 0.101]$, {\bf(c)} $Z = [4.3 - 5]$ and $\gamma = [0.101 - 0.175]$, {\bf(d)} $Z = [5 - 5.175 (SS)]$ and $\gamma > 0.175$, at $\phi = 0.61$. The strain values, for data $1/\gamma_g - 2$, in Fig. $4$(c) of the paper is the strain value at the lower end of the $Z$ window. Data is shown for two values of strain steps ($d\gamma$) used during AQS method. The power law form extends to at least three decades in all the regions. The power law exponent does not vary much with densities for the same window of $Z$ values.}
\end{figure*}

\begin{figure*}[h] 
\includegraphics[scale=0.5,angle=0]{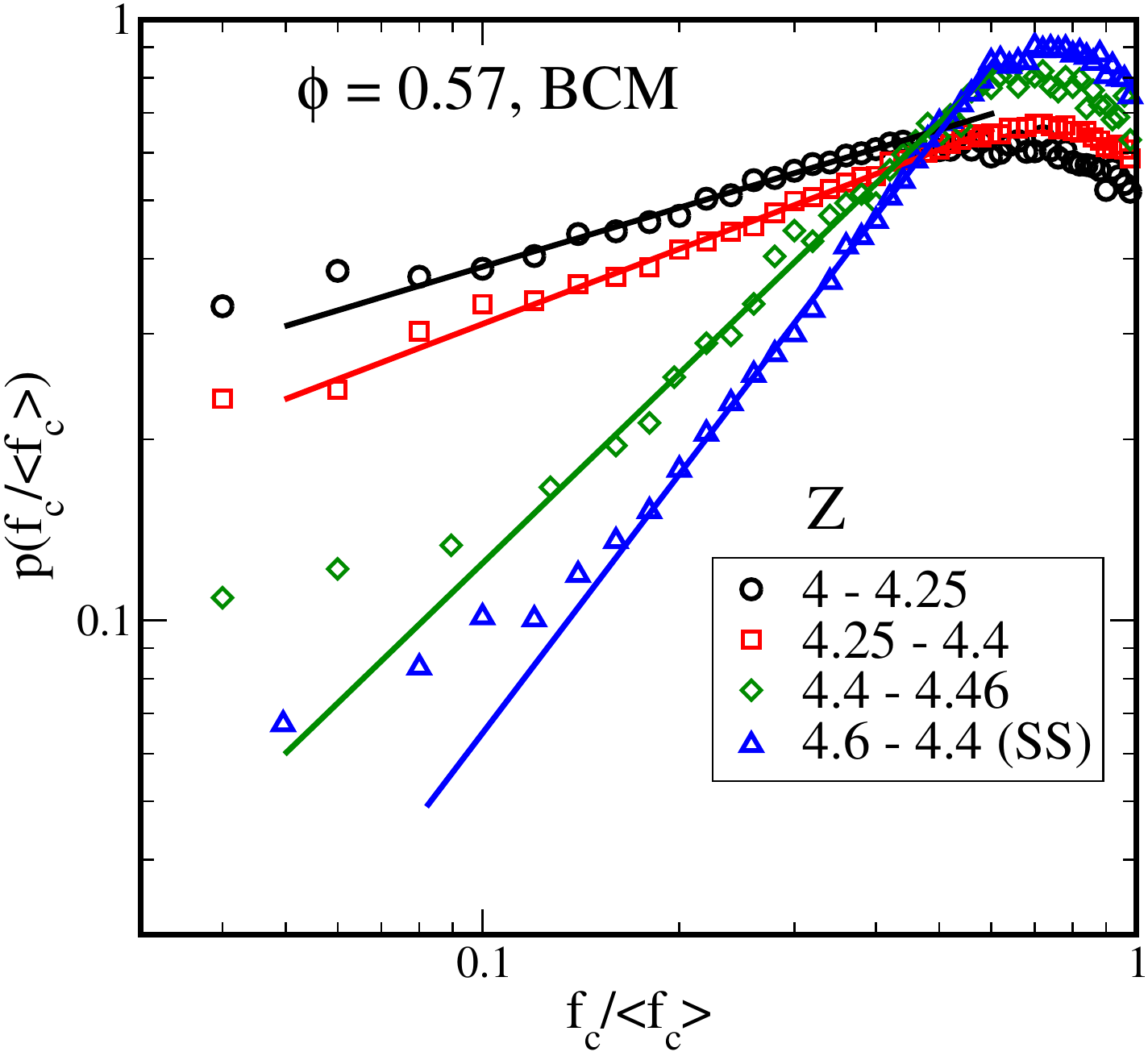}
\includegraphics[scale=0.5,angle=0]{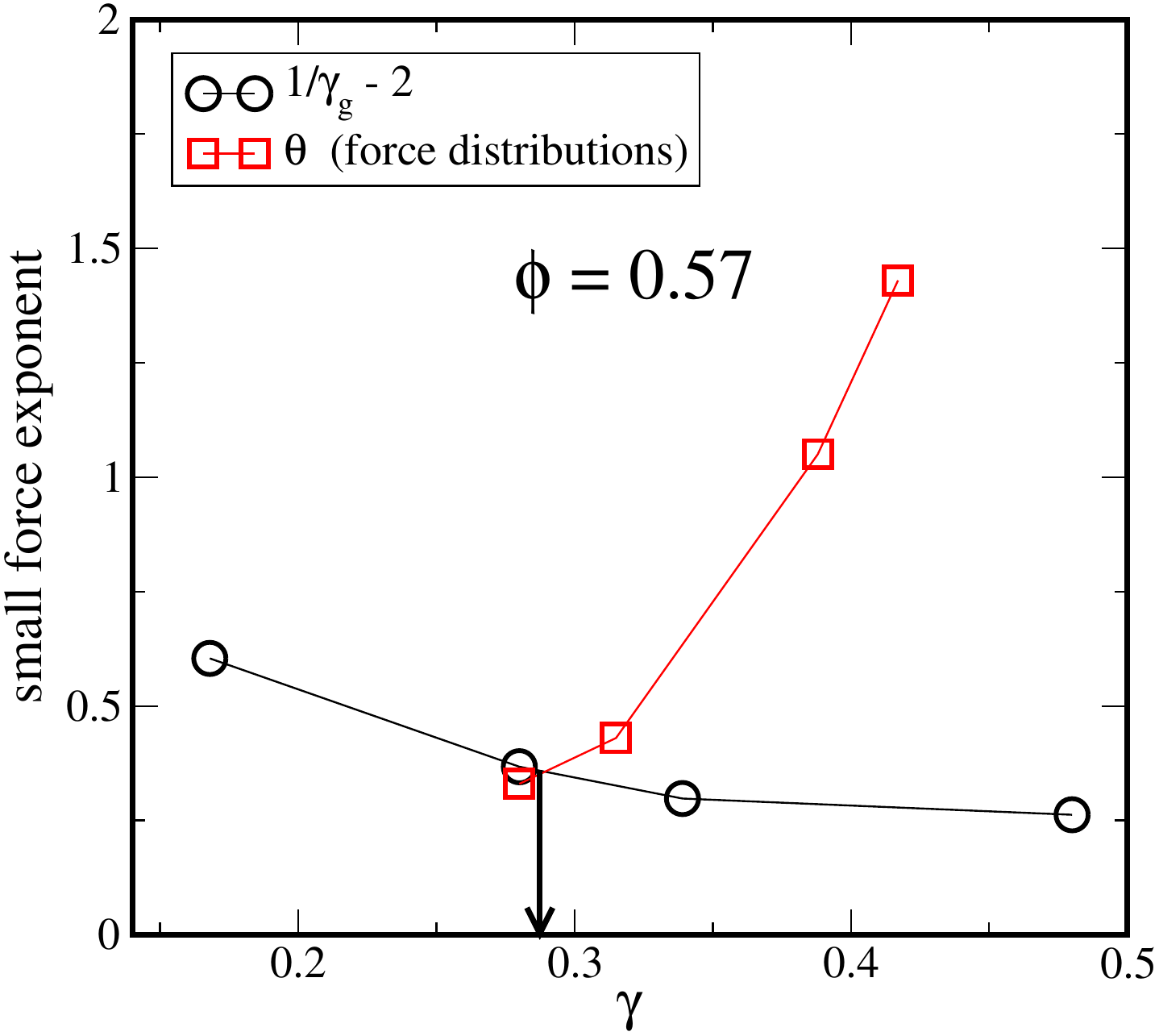}
\caption{\label{figstb057} {\bf (a)}  Small force distribution as a function of strain, shown for $\phi=0.57$, for different windows of $Z$ (or $\gamma$) values. SS implies steady state $Z$ value. {\bf (b)} Exponent values of small force distributions should be than greater than ${\frac{1} { \gamma_g} - 2}$, for stability of the packings. The strain values for small force exponents are the strain value at the lower end of the $Z$ window, obtained from Fig. $2$(a) of the paper. The strain value indicated by the black arrow is where the system becomes marginal.}
\end{figure*}

\begin{figure*}[htbp!] 
\includegraphics[scale=0.5,angle=0]{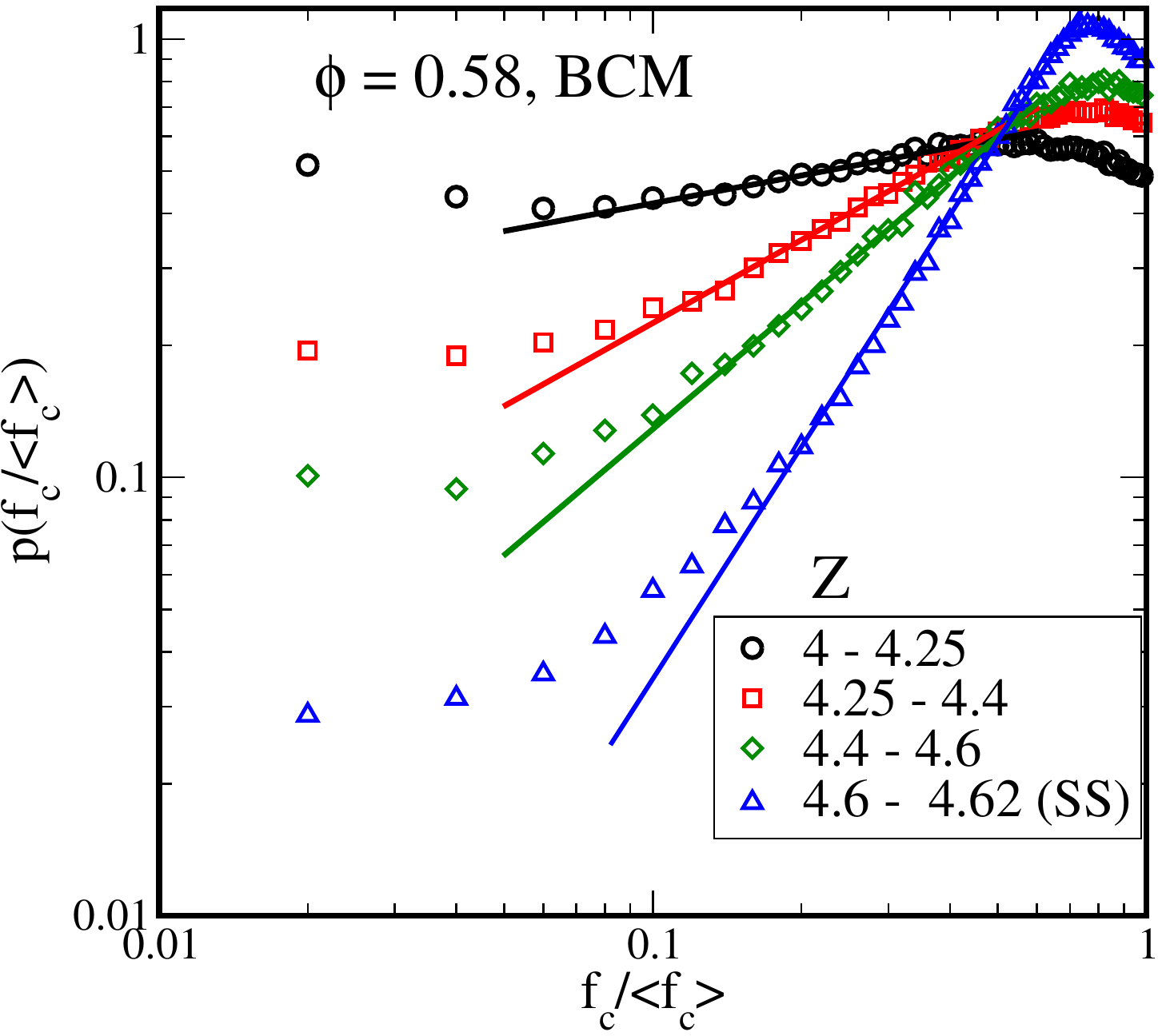}
\includegraphics[scale=0.5,angle=0]{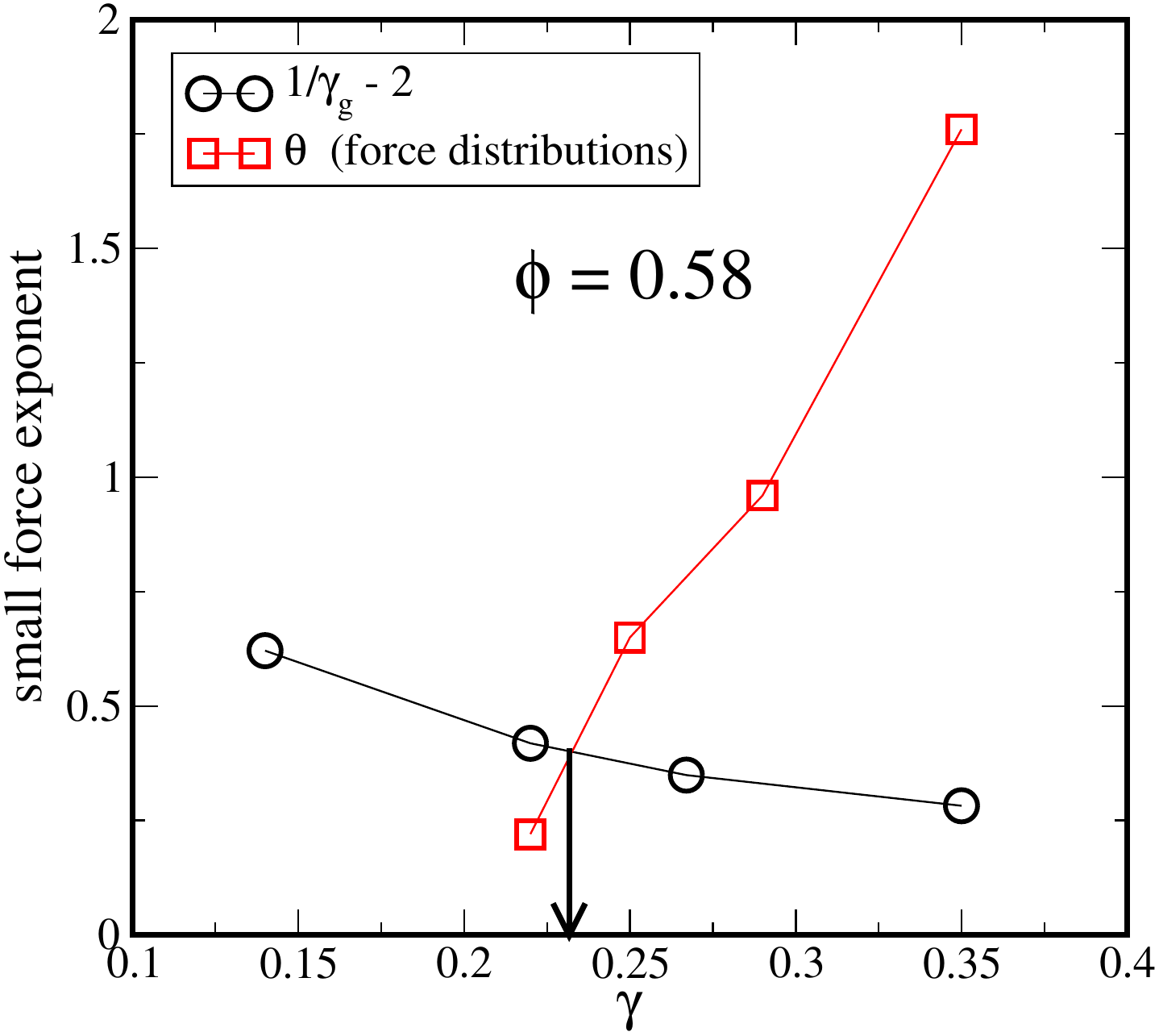}
\caption{\label{figstb058} {\bf (a)}  Small force distribution as a function of strain, shown for $\phi=0.58$, for different windows of $Z$ (or $\gamma$) values. SS implies steady state $Z$ value. {\bf (b)} Exponent values of small force distributions should be than greater than ${\frac{1} { \gamma_g} - 2}$, for stability of the packings. The strain values for small force exponents are the strain value at the lower end of the $Z$ window, obtained from Fig. $2$(a) of the paper. The strain value indicated by the black arrow is where the system becomes marginal.}
\end{figure*}

In Fig. $4$(c) of the paper, we showed force distributions obeying a power law distribution $P(f) \sim f^{\theta}$, as the jamming strain is approached. In Fig. \ref{figgofr}, we show $g(r)$ having a power law form near contact for different $Z$ (or $\gamma$) windows. Using the power law exponents of $g(r)$ ($\gamma_g$), we compute the force distribution exponent required for stability against extended modes. In Fig. \ref{figstb057} and \ref{figstb058}, we show small force distributions and its exponents for $\phi=0.57$ and $\phi=0.58$. As we approach the shear jamming strain, the  system loses its stability. Which also occurs when the low density limit is approached. 

\section{Spatial Correlations of Stress}
\begin{figure*}[htbp!] 
\includegraphics[scale=0.5,angle=0]{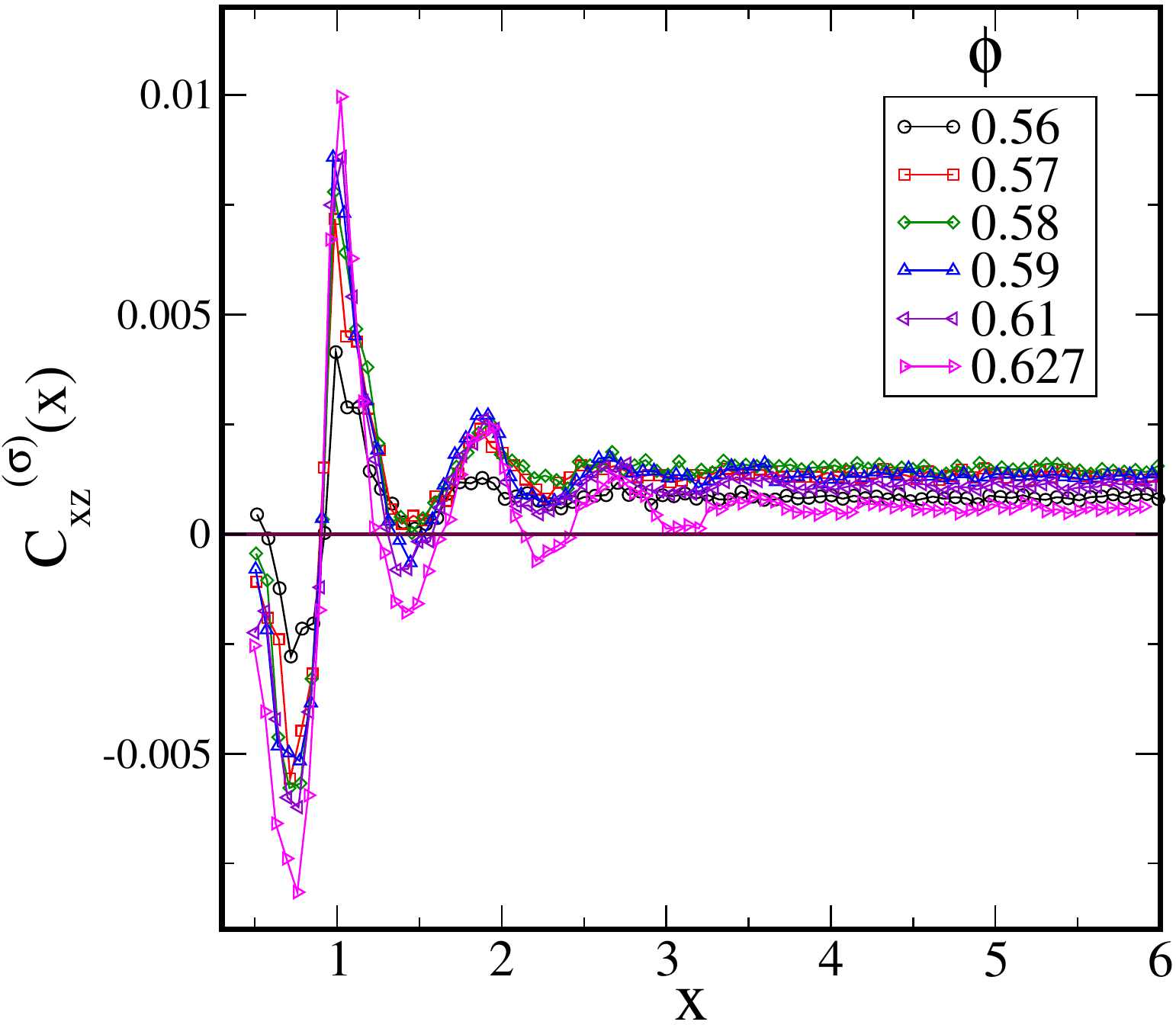}
\includegraphics[scale=0.5,angle=0]{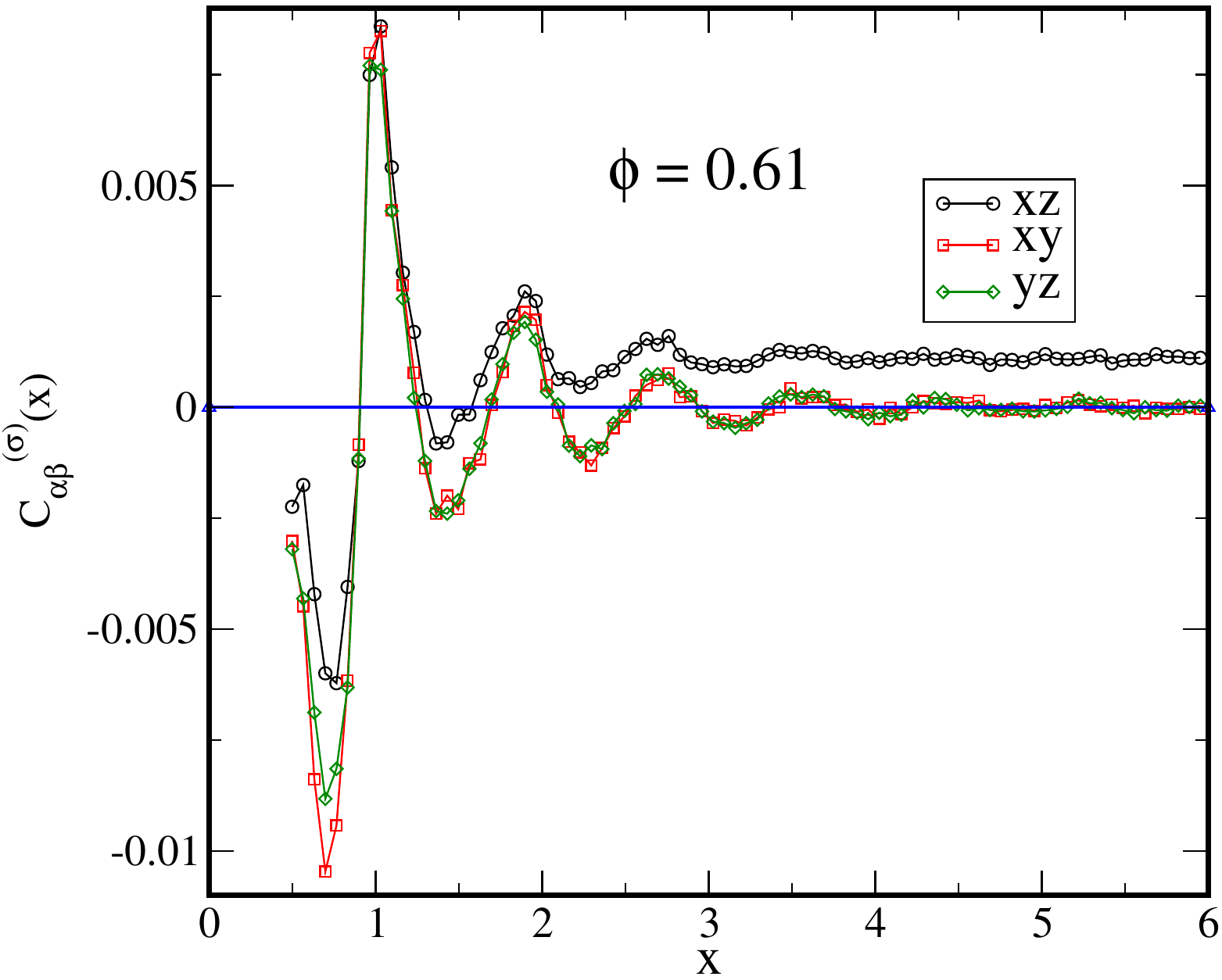}
\caption{\label{figstre} {\bf (a)} Stress-stress spatial correlation as a function of density. {\bf(b)} Spatial contact stress correlation at $\phi = 0.61$ along different planes, $xz$ being the shear plane, in the steady state. Shear jammed packings have non-zero stress correlations along the shear plane and zero in other planes. The data are averaged over $500$ independent solutions for a given contact network and density.}
\end{figure*}

The presence of non-zero shear stress and non-zero spatial stress correlation of bonds in the shear plane, even in the thermodynamic limit, distinguishes shear jammed packings from isotropically jammed packings, \cite{baity2016}. We compute spatial correlation of stresses in the shear plane ($xz$) for the SS configurations at different densities. The correlation is defined as follows:
\begin{equation}
 C_{\alpha\beta}^{(\sigma)} = \textlangle \sum_{k \neq 0} \sigma_{\alpha\beta}^{(0)} \sigma_{\alpha\beta}^{(k)}  \delta([ X^{(0)} - X^{(k)}] - X )  \textrangle
\end{equation}
where $\sigma^{k}$ is the stress on the bond $k$ and $X^{k}$ is the position of bond $k$. Along the shear plane $xz$, we observe that the correlation are finite and is zero in the other planes.
Along the shear plane $xz$, we observe that at large $X$ the correlations are finite for densities above $\phi=0.55$, see Fig. \ref{figstre}(a). In Fig. \ref{figstre}(b), we show spatial stress correlation of bonds along different planes for $\phi=0.61$, which are finite at large $X$ for the shear plane $xz$ and zero in the other planes. 

\section{Density-Strain Phase Diagram for the Two Dimensional System}
\begin{figure*}[h]
\includegraphics[scale=0.55,angle=0]{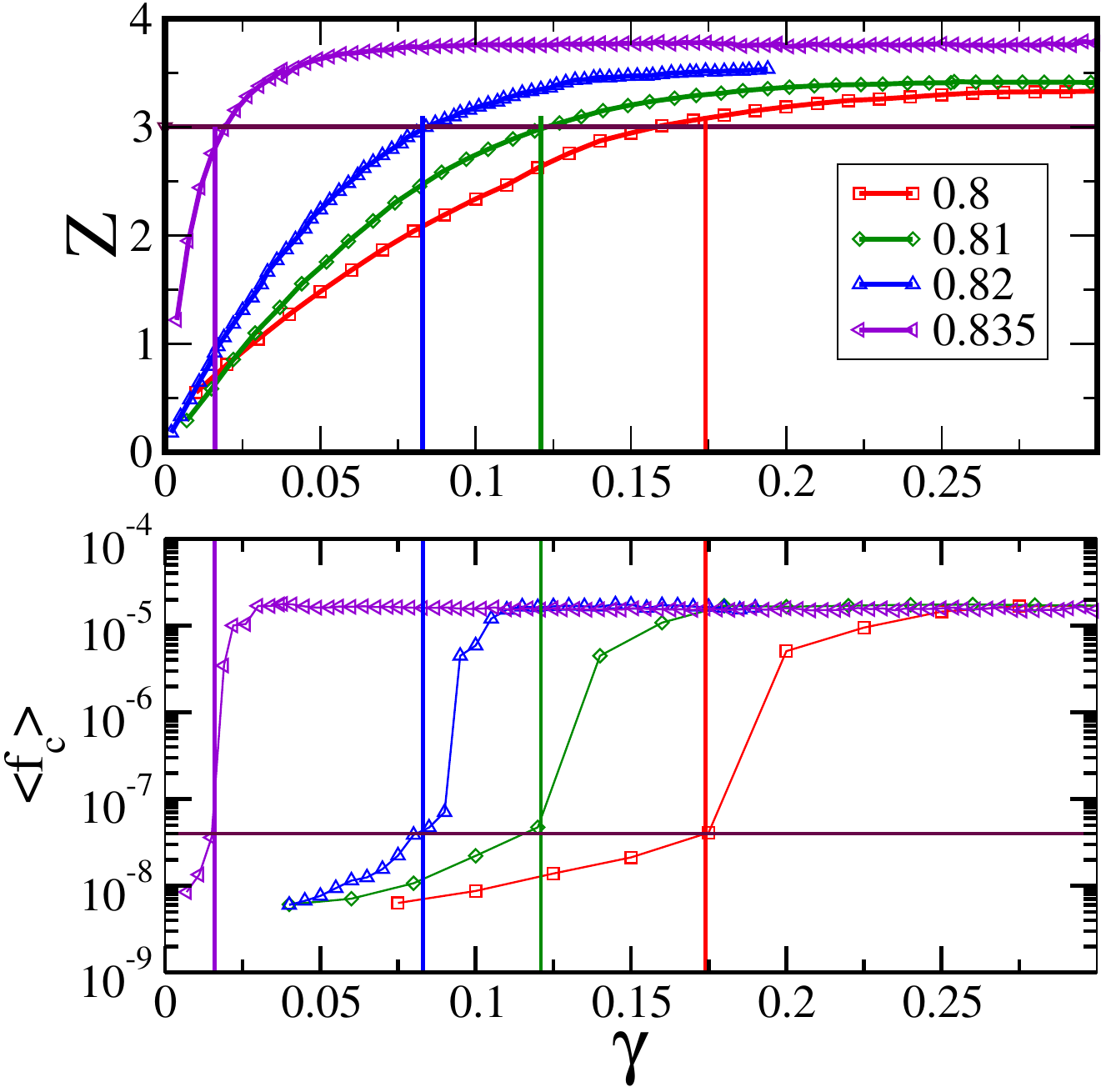}
\includegraphics[scale=0.55,angle=0]{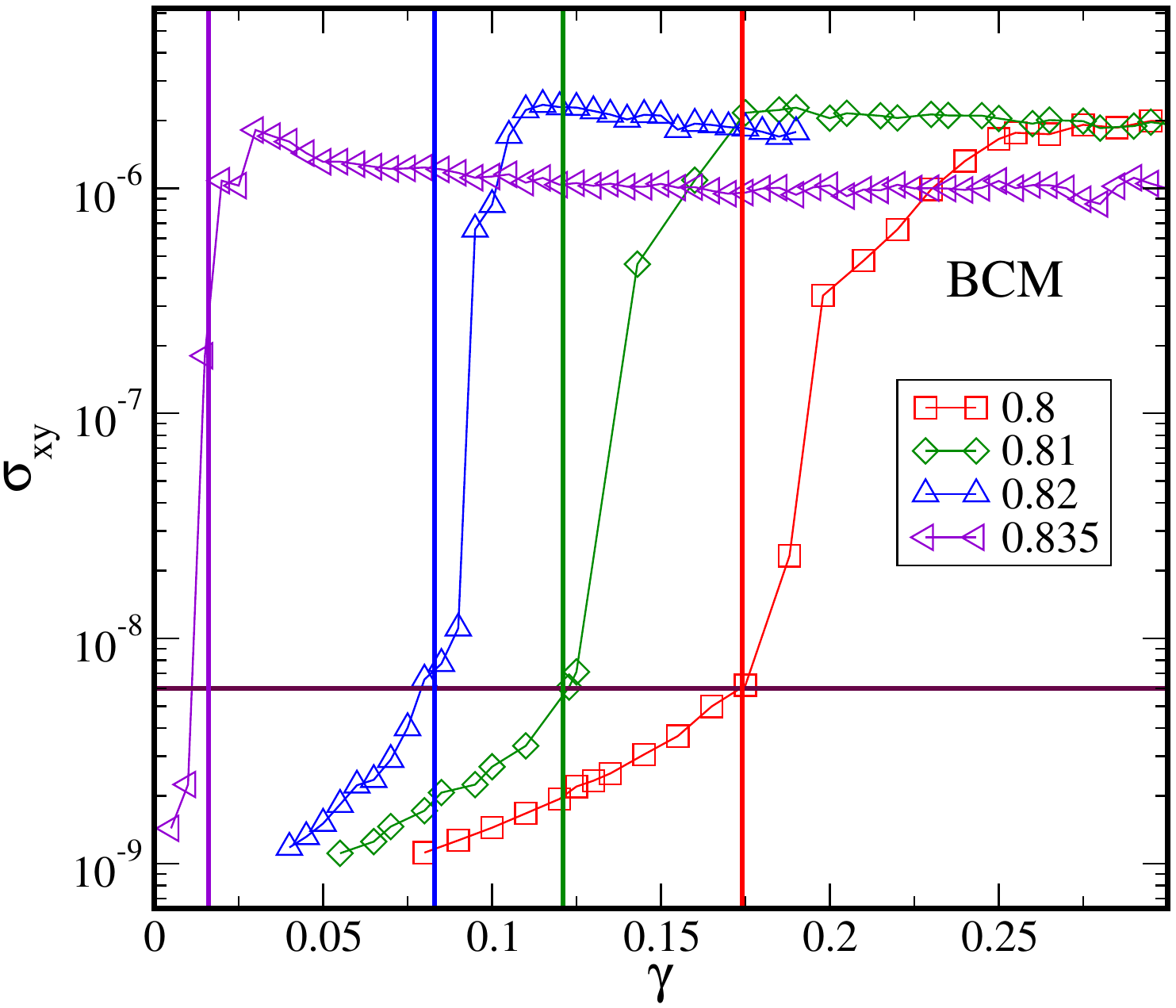}
\caption{\label{fig5}  Mean contact number $Z$, contact force $\langle f_c \rangle$ and  stress as a function of strain for the infinite friction case for the two dimensional packings. $\langle f_c \rangle$ shows a jump at the shear jamming transition. The contact forces are computed from the force solutions obtained using the BCM method. The maroon horizontal line in the $Z$ {\it vs.} $\gamma$ plot marks $Z=3$, and the horizontal lines in the stress and $\langle f_c \rangle$ plots marks the cutoff used to identify the shear jamming strain.  The vertical lines marks shear jamming (SJ) strain values corresponding to sudden increase in  $\langle f_c \rangle$  value, shown in Fig. $5(a)$ of the paper. The shear jamming strain values identified using  $\langle f_c \rangle$  also marks the strain at which the stress shows a sudden jump.}
\end{figure*}

\begin{figure*}[h]
\includegraphics[scale=0.42,angle=0]{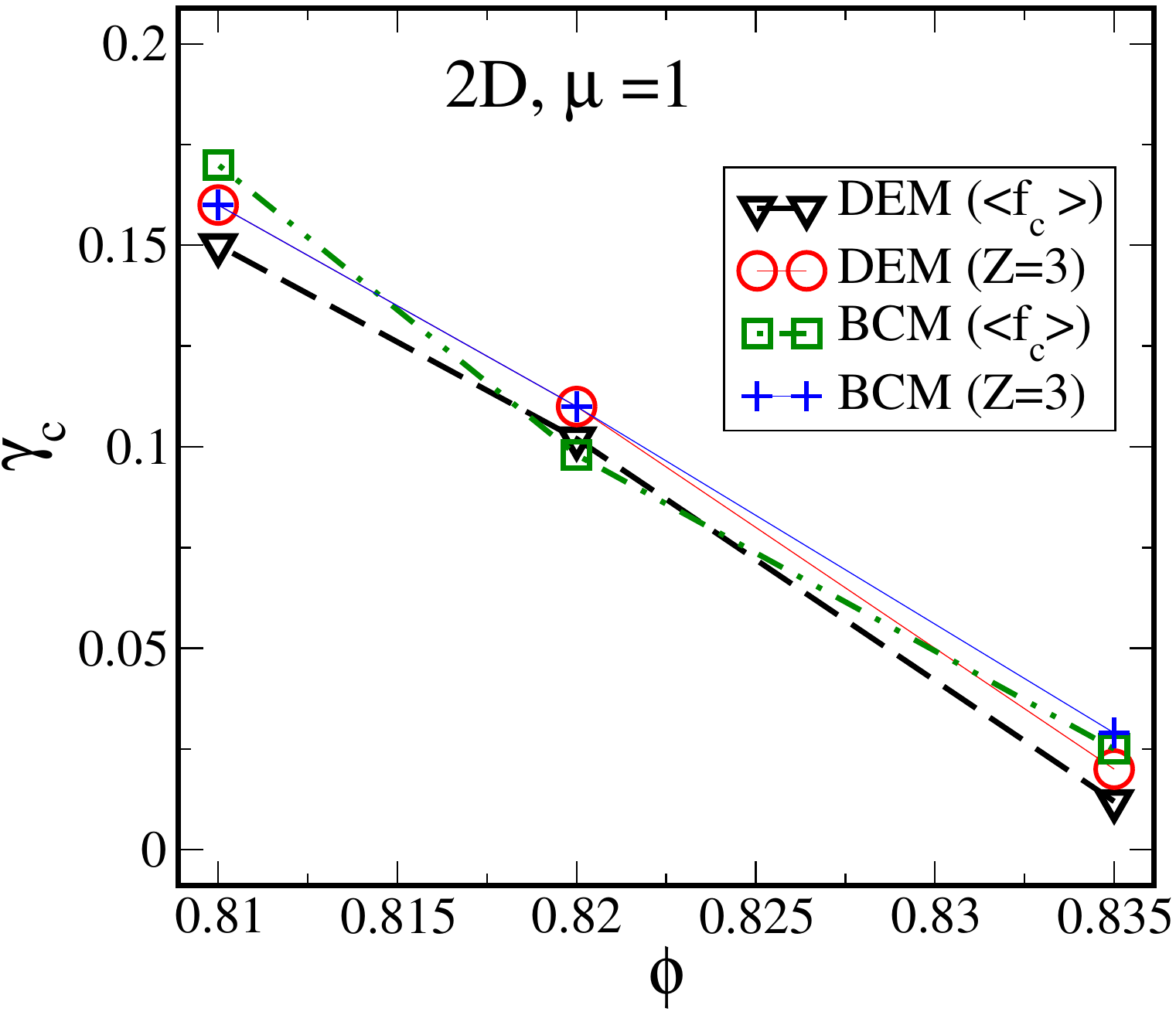}
\includegraphics[scale=0.38,angle=0]{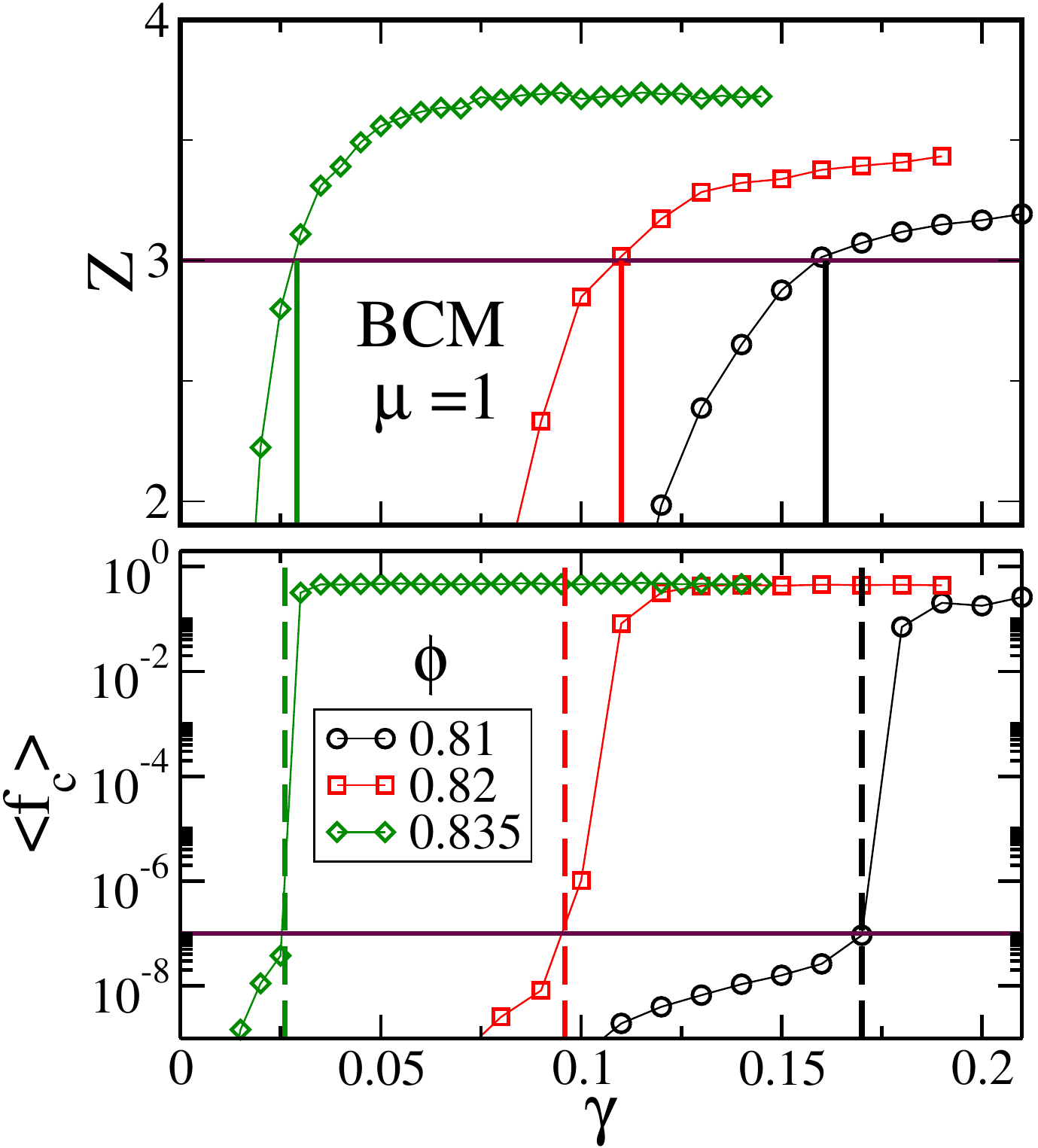}
\includegraphics[scale=0.38,angle=0]{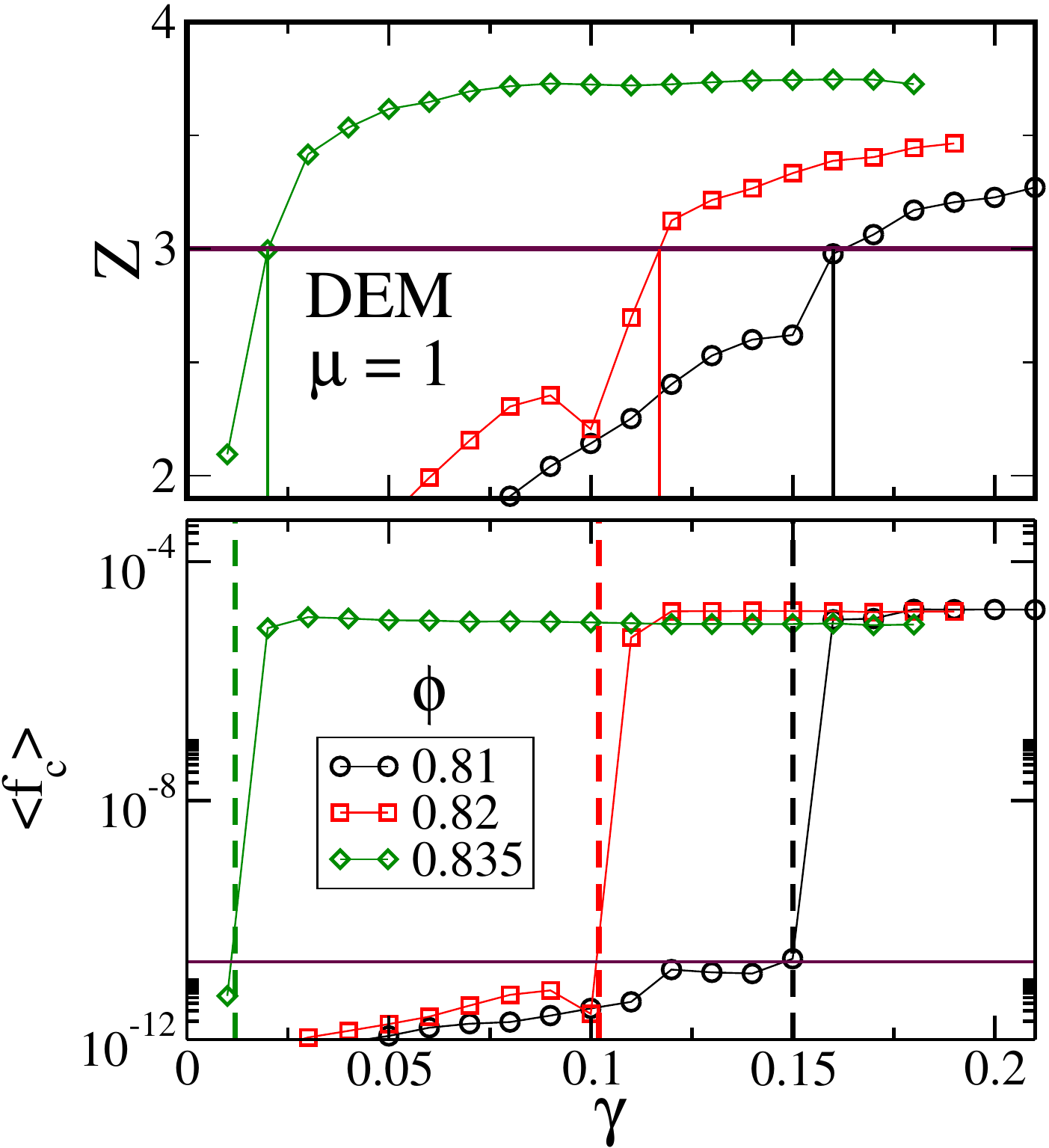}
\caption{\label{fig6} {\bf(a)} Jamming phase diagram for the two dimensional system at finite friction $\mu = 1$, which indicates that $\phi=0.81$ is the lower density limit for the physical value of friction coefficient of $\mu=1$. $Z$ and $\langle f_c \rangle$ as a function of strain, shown for different densities, for {\bf (b)} BCM ($\mu=1$) and {\bf (c)} DEM ($\mu=1$) protocols. The shear jamming strain values obtained from BCM and DEM methods are close to each other and closely corresponds to $Z=3(=D+1)$. The vertical lines (dashed) shows strain values corresponding to the shear jamming (SJ) transition, $\langle f_c \rangle$ show discontinuous jumps, and the bold vertical lines correspond to $Z = 4  (= D+1)$.}
\end{figure*}

In this section, supporting plots for the 2D phase diagram discussed in the main text. In Fig. \ref{fig5}, we show supporting data for the BCM results shown in Fig. $5(a)$ of the paper.  The average contact force and stress obtained from the BCM method are shown. The data is averaged over $5$ configurations.  For $\phi=0.8$ average is done over $15$ configurations. The BCM jamming strain is obtained from the strain values marked by the vertical lines in plot of $<f_c>$, which is very close to $Z=D+1$. In Fig. \ref{fig6}, contact force data for the finite friction coefficient $\mu=1$ from BCM is shown along with DEM data for $\mu=1$, showing that the jamming strain values for BCM and DEM are close to each other and to $Z = 3 (= D+1)$.

\section{Rigid and Rigid+Stress Percolation in Two Dimensions}
\begin{figure*}[htbp!] 
\includegraphics[scale=0.55,angle=0]{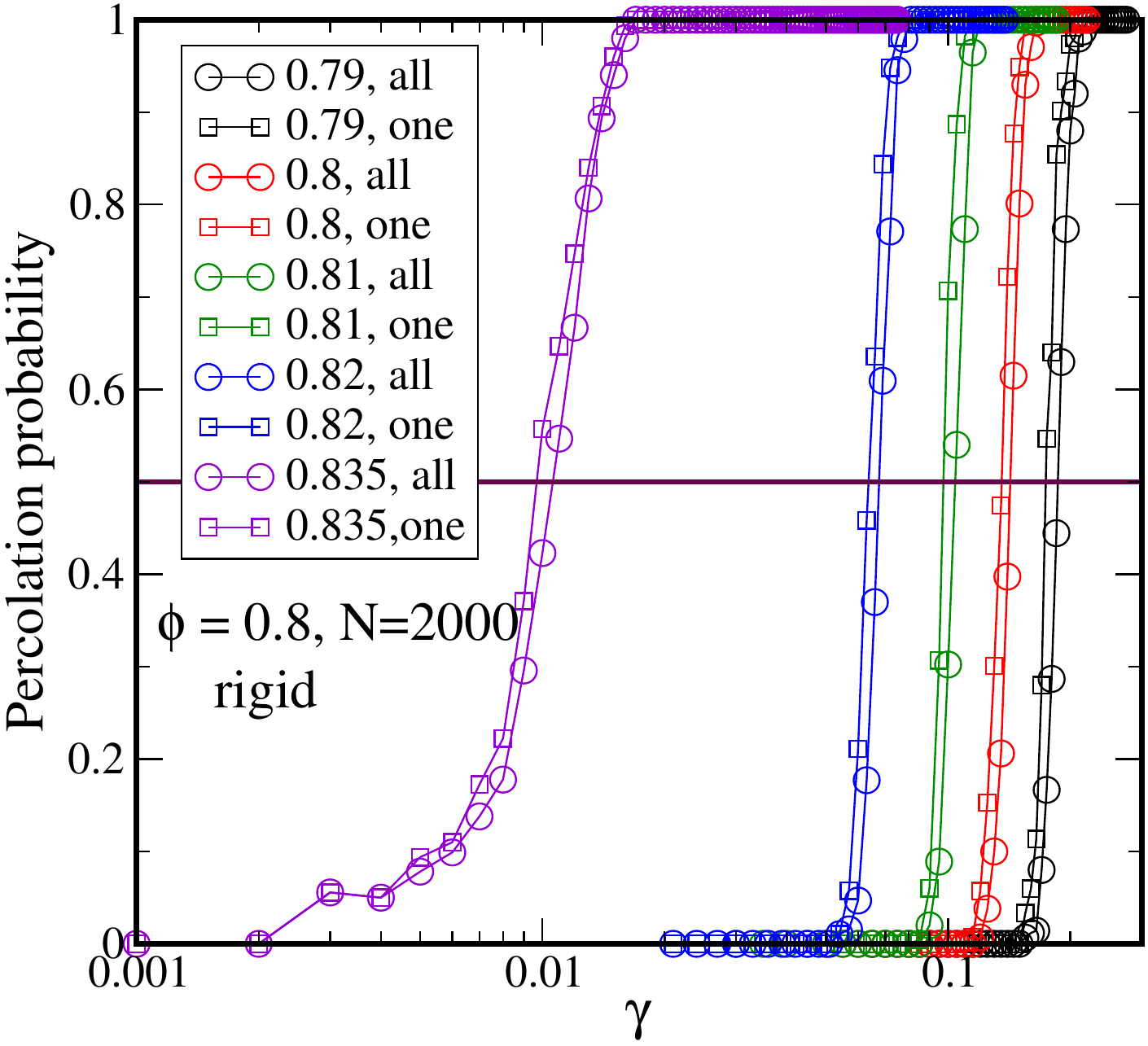}
\includegraphics[scale=0.55,angle=0]{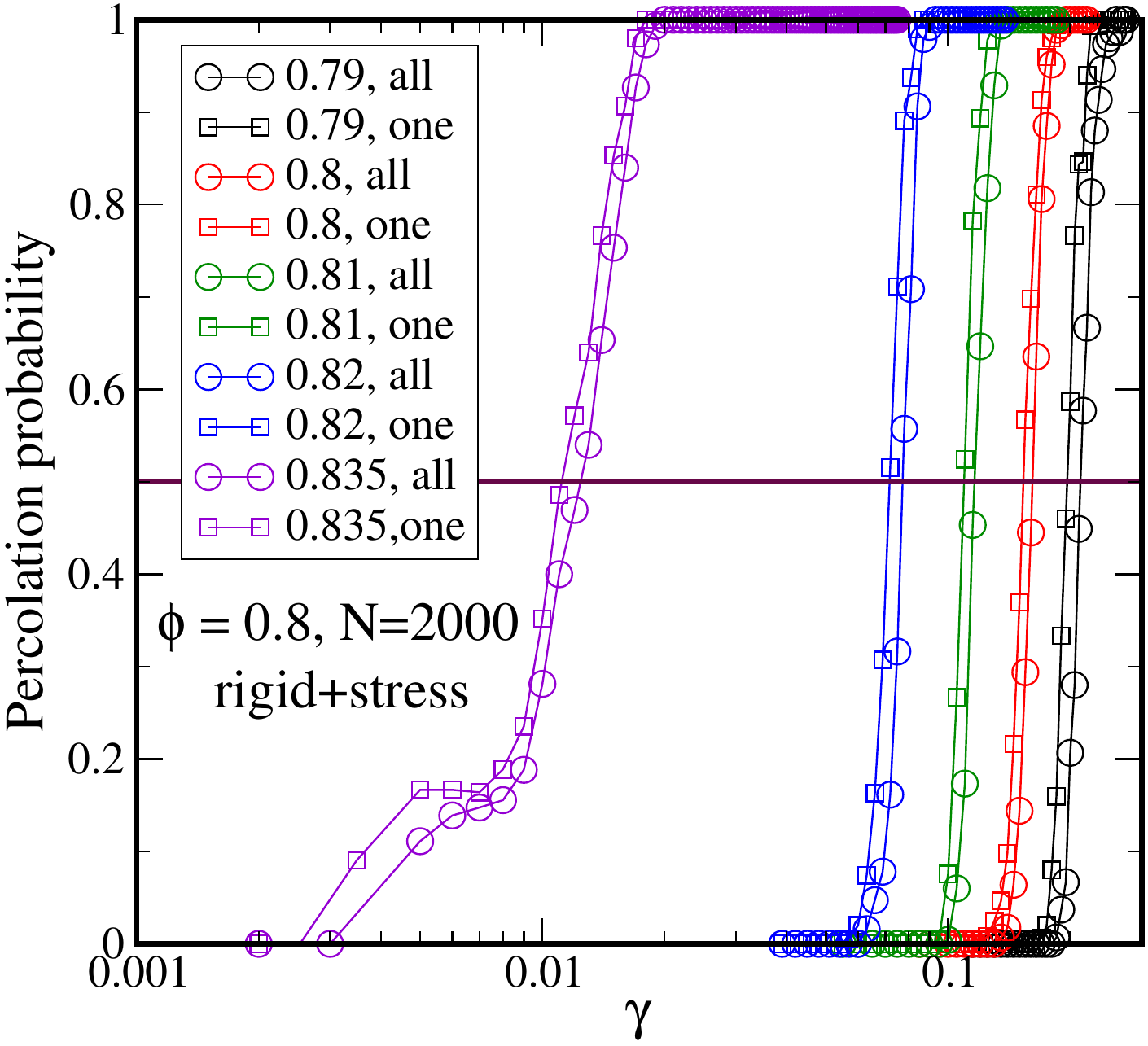}
\caption{\label{fig3} {\bf(a)} Percolation probability of rigid clusters as a function of strain $\gamma$ for different densities for the two dimensional system for $N = 2000$. {\bf(b)} Percolation probability of over-constrained regions (rigid+stress) as a function of strain $\gamma$ for different densities. The label {\it one} refers to the case where the percolation probability is computing considering spanning cluster along one direction, whereas {\it all} refers to the case where the percolation probability is  computed requiring that a spanning cluster is system spanning in both the $x$ and $y$ directions. 
The rigid+stress percolation strain values in Fig. $5$(a) of the paper correspond to a percolation probability of $0.5$ (maroon horizontal line) of the {\it all} case.}
\end{figure*}

\begin{figure*}[htbp!] 
\includegraphics[scale=0.4,angle=0]{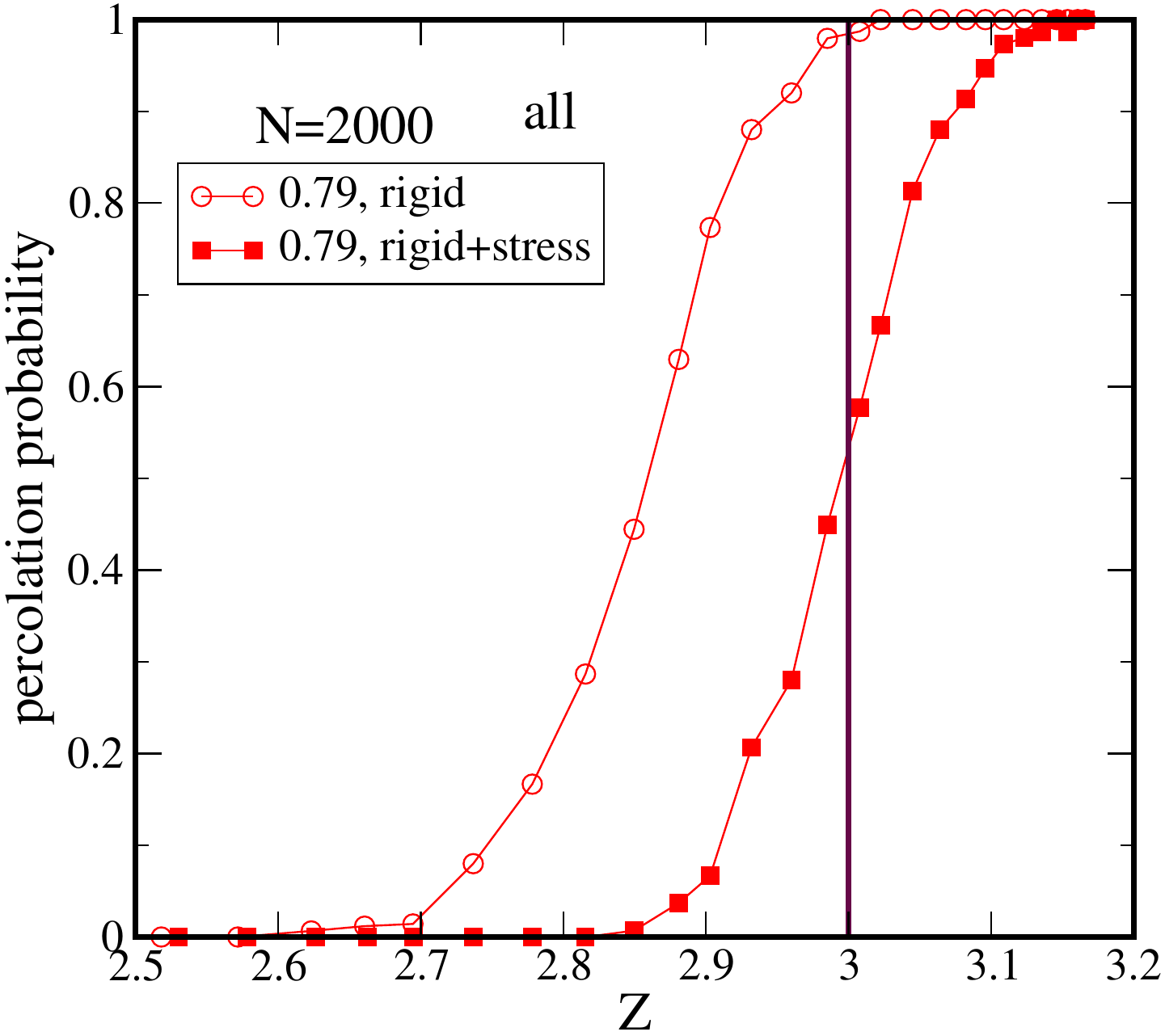}
\includegraphics[scale=0.4,angle=0]{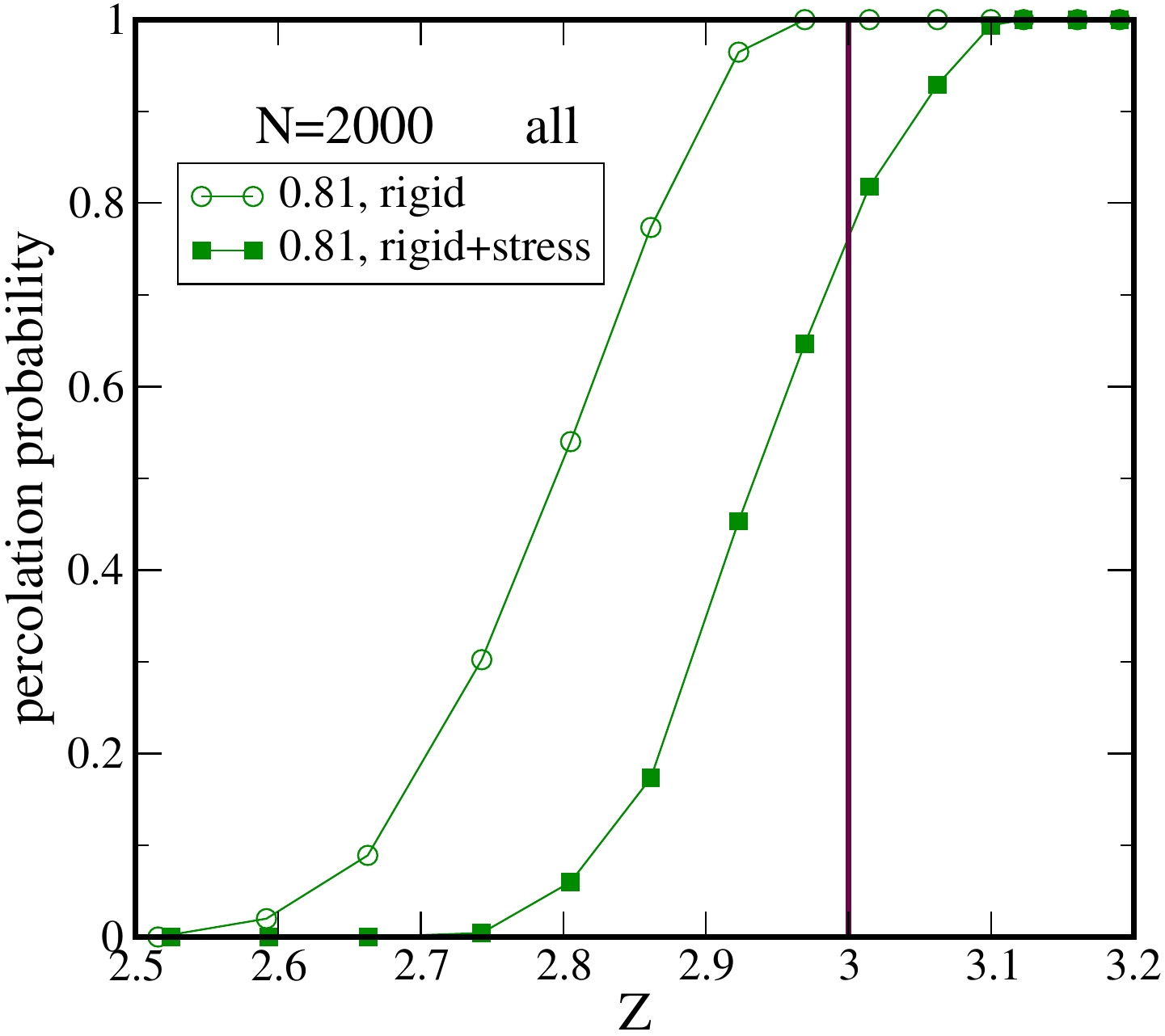}
\includegraphics[scale=0.4,angle=0]{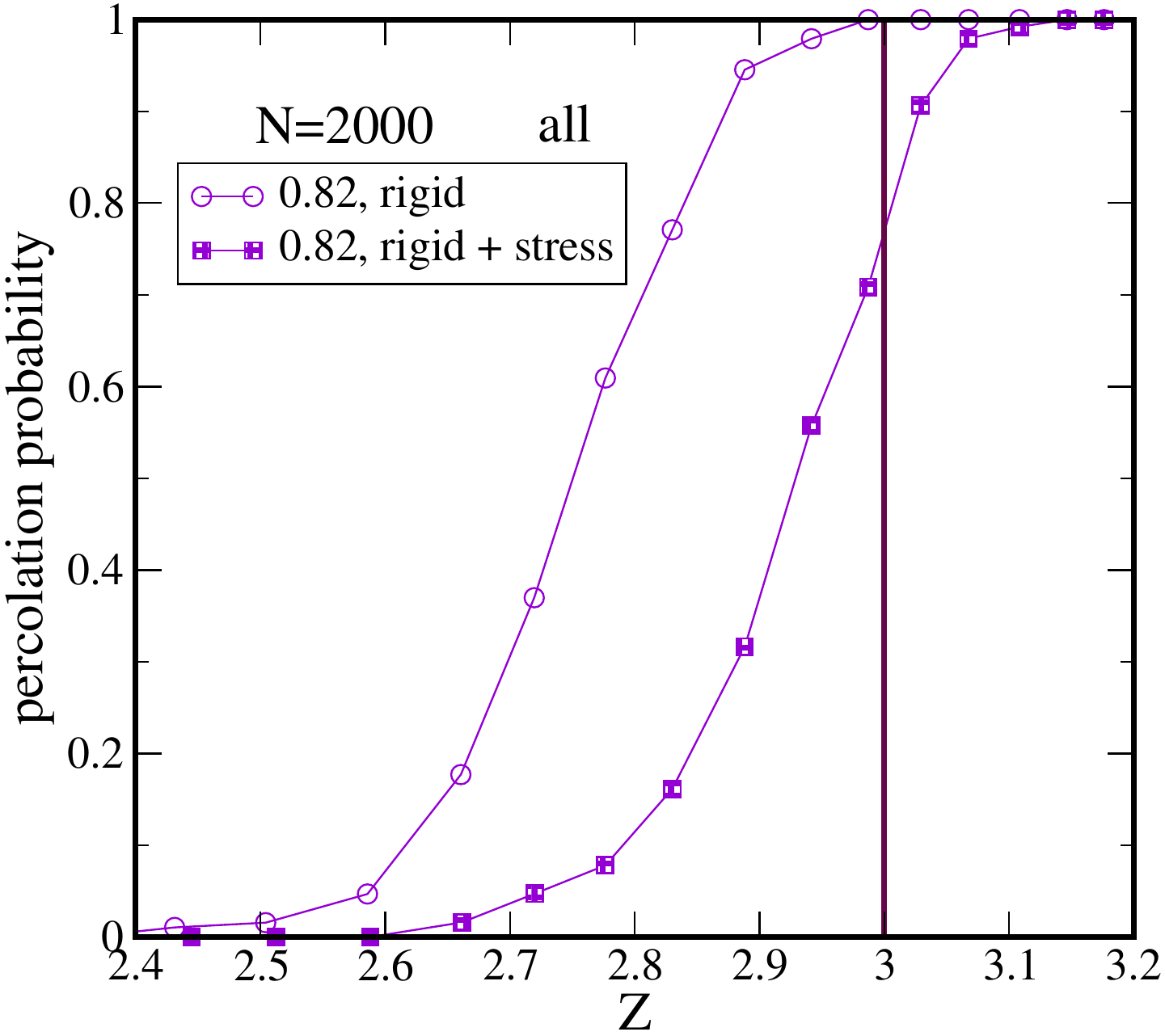}
\caption{\label{fig4} Percolation probability as a function of $Z$, shown for {\bf (a)} $\phi=0.79$, {\bf (b)} $\phi=0.81$ and {\bf (c)} $\phi=0.82$ for the two dimensional system, $N=2000$. The data shown is for the {\it all} case, where percolation probability is computed requiring a spanning cluster along both the directions. These plots show the presence of an intermediate phase for all densities $\phi \geq \phi_{RLP}=0.79$}
\end{figure*}

\begin{figure*}[h]
\centering 
\includegraphics[scale=0.55,angle=0]{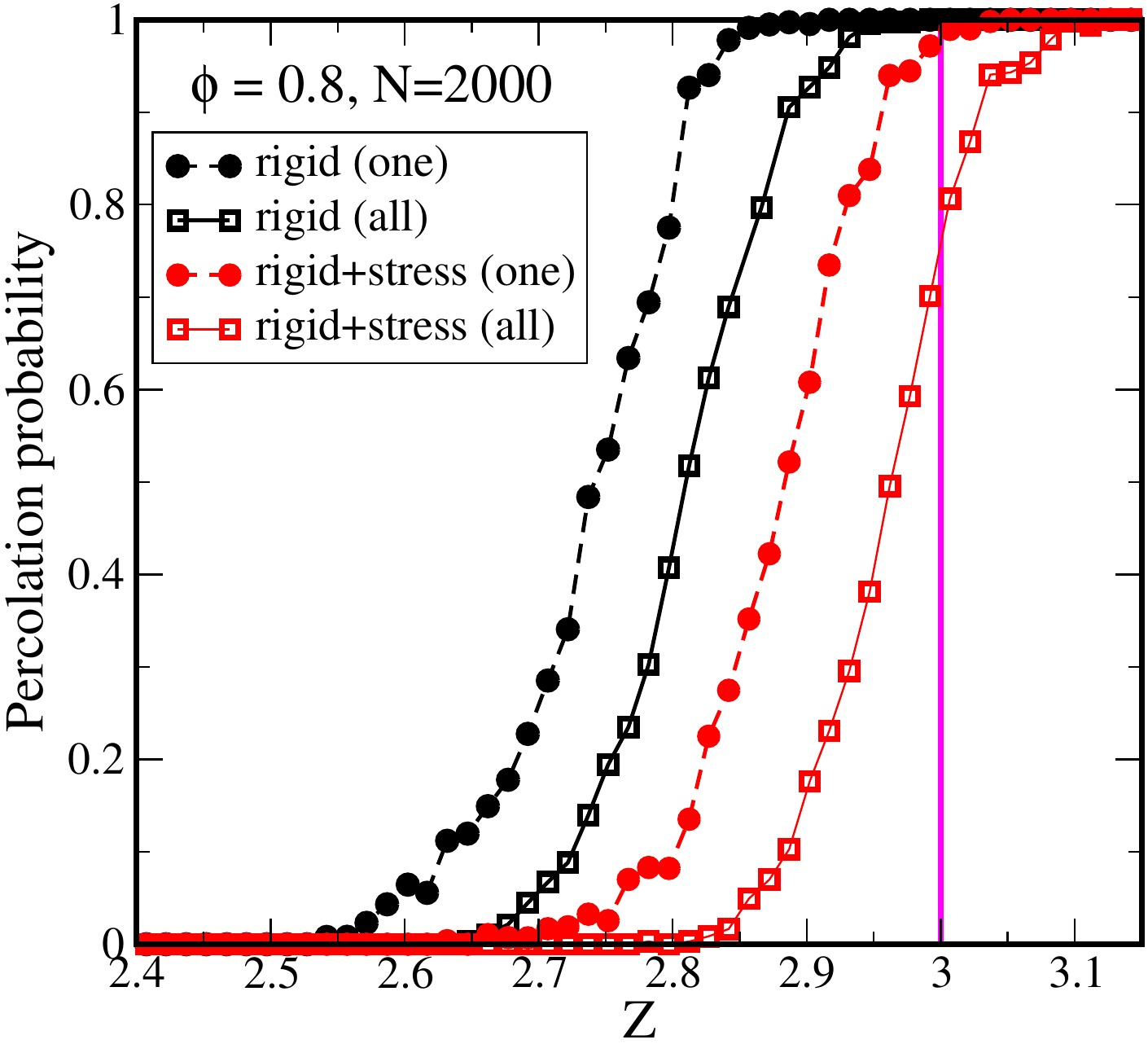}
\caption{\label{fig8} Rigid percolation and rigid +stress percolation data along {\it one} and {\it all} ($x$ and $y$) directions occurs at different $Z$ values. The percolation of rigid+stress clusters that percolate along one direction can support stress transmission along the direction of percolation, which indicates the presence of fragile force networks in the region between rigid percolation and rigid+stress percolation or shear jamming transition.}
\end{figure*}

In Fig. \ref{fig3}, we show rigid and over-constrained (rigid+stress) percolation probabilities as a function of strain, for $N=2000$. The rigid and rigid+stress strain values in Fig. $5$(a) of the paper corresponds to percolation probability of $0.5$ of case {\it all} (where we require percolation to occur along both axes). The percolation along {\it one} and {\it all} directions occurs at different strain values. This separation is more apparent as a function of $Z$ rather than $\gamma$, as shown in Fig. \ref{fig8}. To show the presence of an intermediate phase for different densities, we show the percolation of rigid clusters and over-constrained regions (rigid+stress) as a function of $Z$ (for the {\it all} case) in Fig. \ref{fig4}. Since the configurations are anisotropic, the rigid and rigid+stress percolations along {\it one} and {\it all} directions are well separated, as shown in Fig. \ref{fig8}.  The percolation of rigid+stress clusters that percolate along one direction can support stress transmission along the direction of percolation. Propagation of stress along one direction is characteristic of the fragile force networks.

\section{Intermediate phase in shear jamming}
\begin{figure*}[htbp!]
\includegraphics[scale=0.55,angle=0]{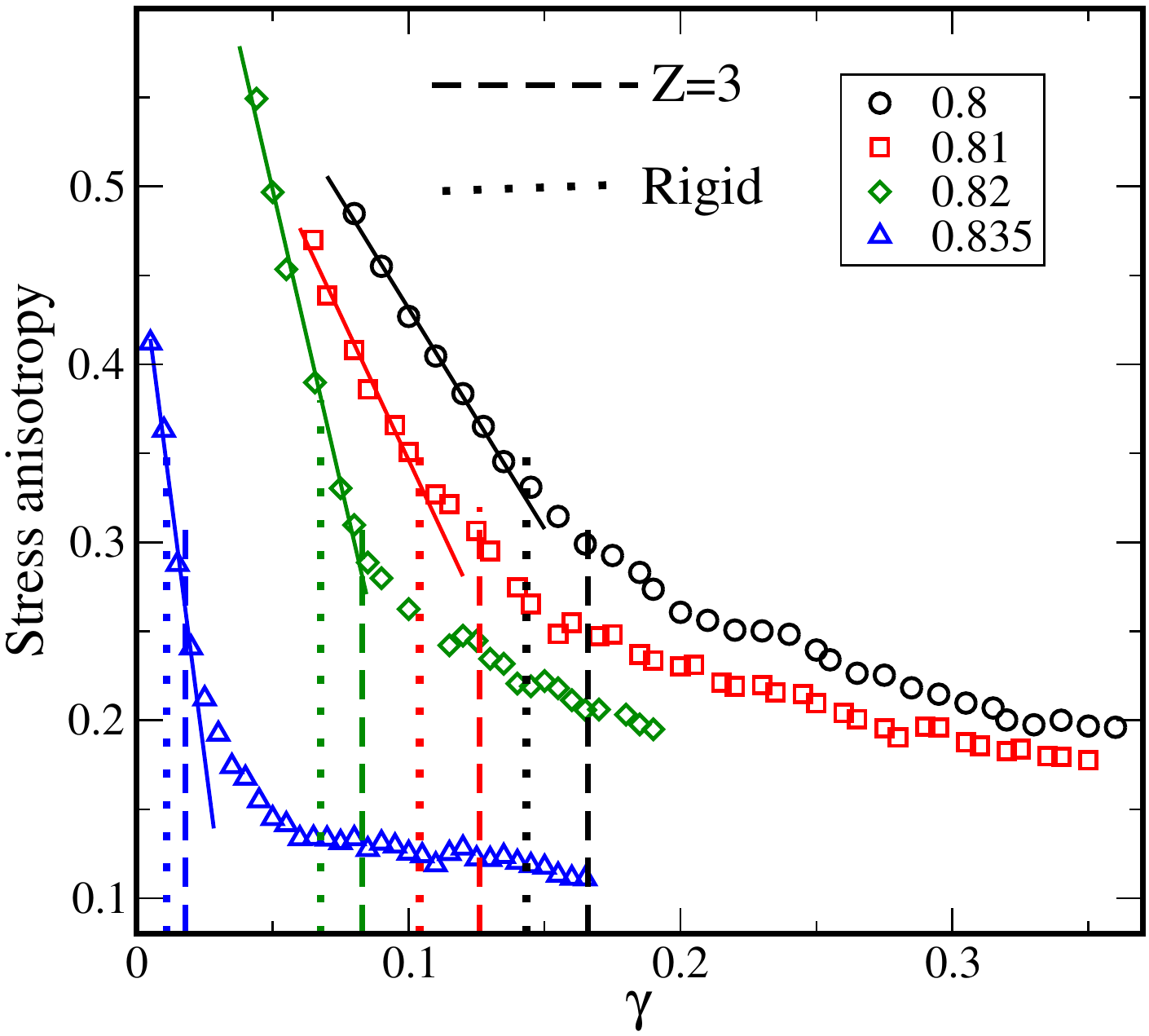}
\includegraphics[scale=0.55,angle=0]{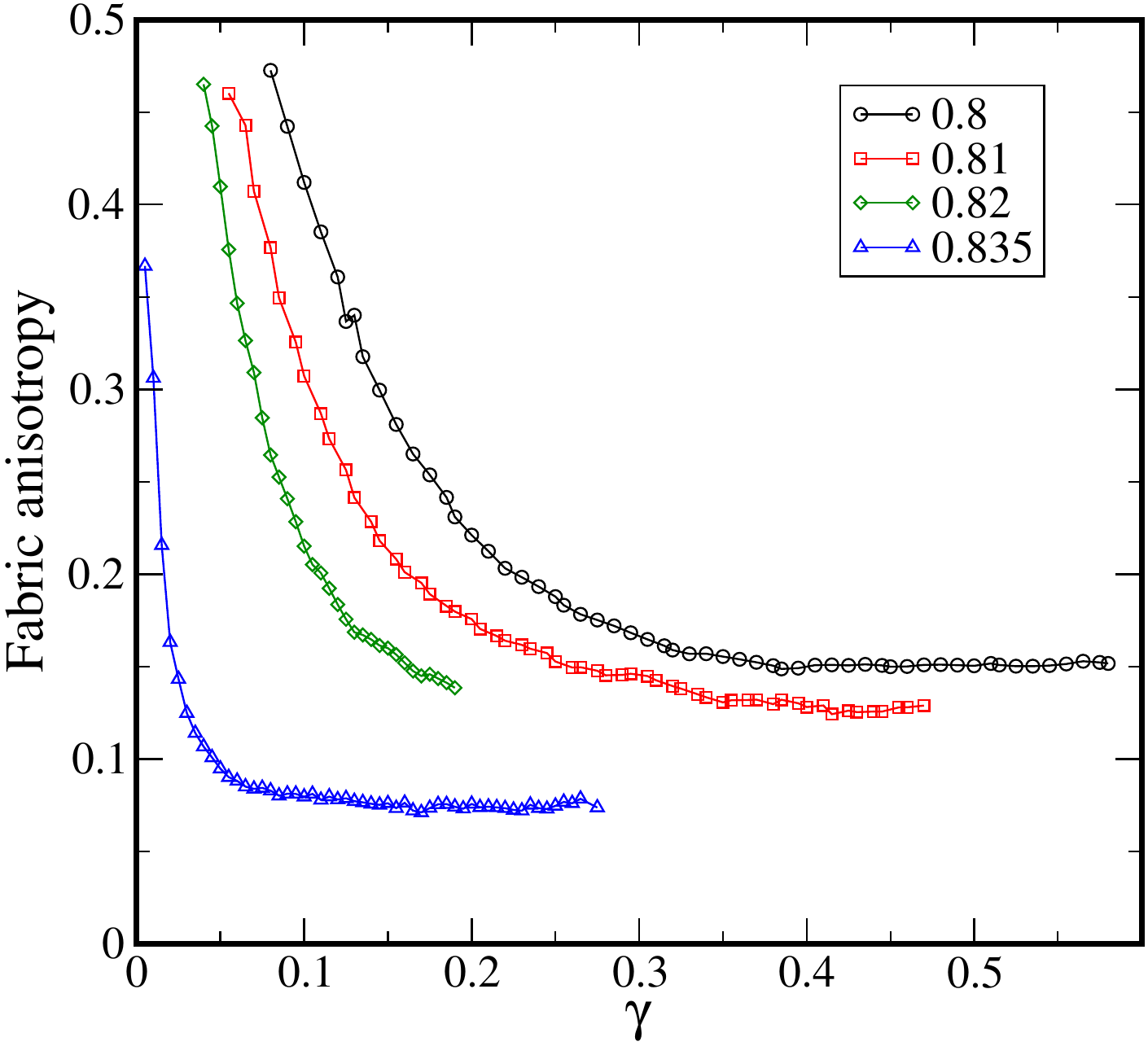}
\caption{\label{fig7} {\bf(a)} Stress anisotropy (SA) and {\bf(b)} fabric anisotropy (FA) as a function of strain, shown for different densities. The force balance solutions are obtained from BCM method in the limit of infinite friction. The stress anisotropy plot shows that in the strain window of rigid percolation and $Z=3(D+1)$ the stress anisotropy begins to saturate, which suggests the presence of fragile force networks in this region or strain window.}
\end{figure*}

 The possibility of fragile force networks to form in the intermediate phase is supported by two results below. In Fig. \ref{fig8}, we show rigid+stress percolation along {\it one} and {\it all} directions to occur at different values of $Z$ and $\gamma$ and the rigid+stress percolation along one direction occurs after the rigid percolation. In the intermediate window, the percolating clusters of over-constrained regions can support strong forces along one direction, which is the definition of fragile force network \cite{cates1998,bi2011}.
We characterize the anisotropy in the sheared two dimensional packings using fabric and stress anisotropy, defined as follows:
\begin{eqnarray}
 \hat{R} = \frac{1}{N}\Sigma_{i \neq j} \frac{\vec{r_{ij}}}{\mid r_{ij}\mid} \otimes \frac{\vec{r_{ij}}}{\mid r_{ij}\mid}  \\
 \hat{\sigma} = \frac{1}{V}\Sigma_{i \neq j} \vec{r_{ij}} \otimes \vec{f_{ij}},
\end{eqnarray}
where $\vec{r_{ij}}$ is the distance between the centers of pair $(ij)$. We diagonalize the fabric (stress) tensor and obtain principal components of fabric (stress) tensor and the corresponding eigenvalues. Eigenvalues of the fabric tensor are $C_1 > C_2$ and the stress tensor are $P_1 > P_2$. $C_1$ ($P_1$) is along the compressive direction and $C_2$($P_2$)is along dilative direction. The stress and fabric anisotropy are defined as follows: 
\begin{eqnarray}
 FA = (C_2 - C_1)/(C_1 + C_2) \\
 SA = (P_2 - P_1)/(P_1+ P_2)
\end{eqnarray}
In Fig. \ref{fig7}, we show the stress anisotropy and fabric anisotropy as a function of strain for different densities. In the stress anisotropy plot, we mark the strain values corresponding to $Z=3$ and rigid percolation. We see that in this window of strain values the anisotropy of the force network starts to deviate from the linear behavior and  to saturate, which supports the above idea. These ideas are speculative at the moment and require further investigation to validate them.

\pagebreak
\section{Evolution of Rigid, Rigid+Stress, and Force Networks}
\begin{figure*}[h]
\includegraphics[scale=0.3,angle=0]{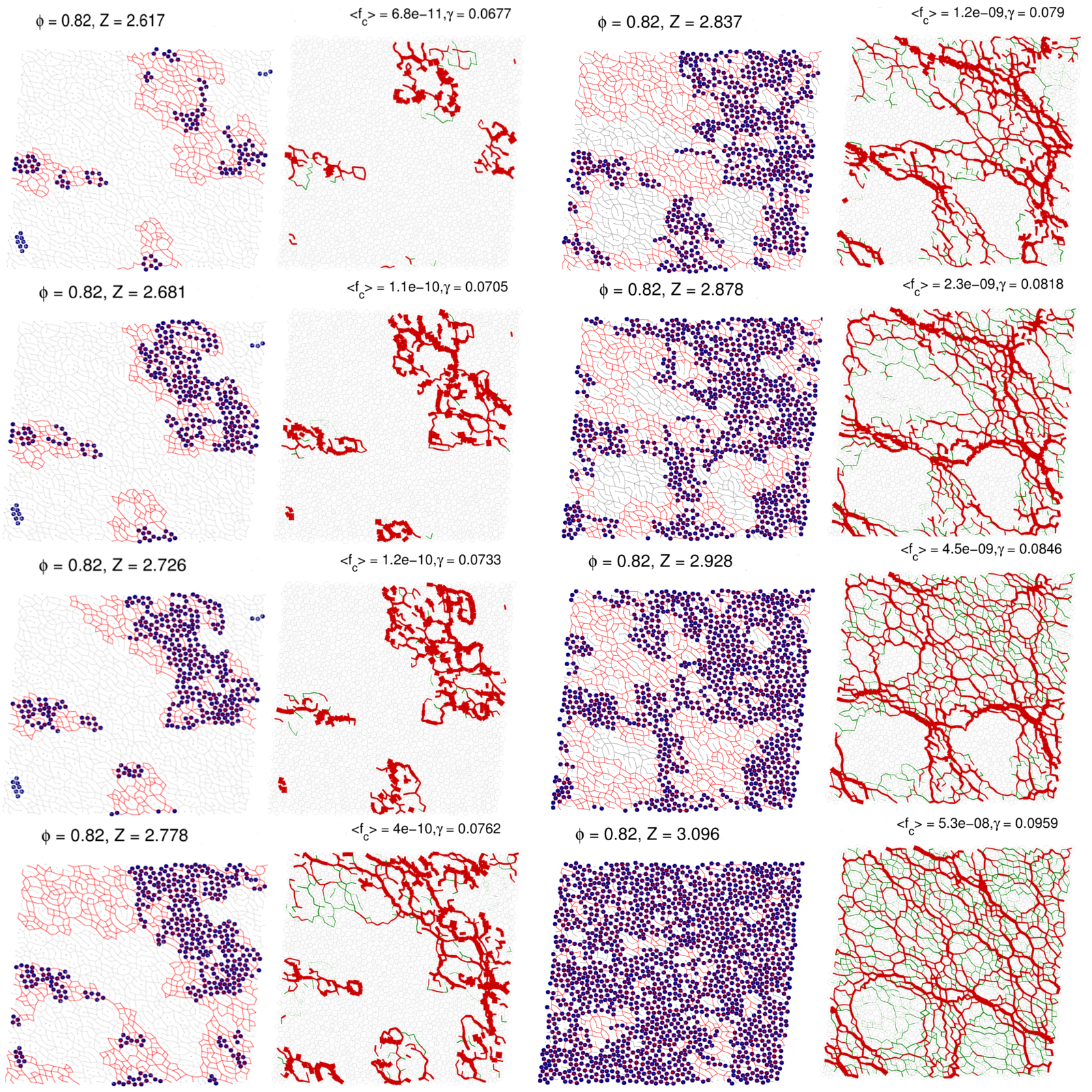}
\caption{\label{fig9} 
The left panels show the evolution of rigid and over constrained (rigid+stress) clusters as a function of strain (red network - biggest rigid cluster, other rigid clusters are not shown, blue filled circles - over-constrained discs and grey network - floppy regions and other smaller rigid clusters). The right panels show the evolution of contact forces $\langle f_c \rangle$ as a function of strain ($f_c > \langle f_c \rangle$ - red bonds, $f_c < \langle f_c \rangle$ - green bonds and $f_c < 10^{-14}$ - grey bonds). Data shown are for $\phi = 0.82$. The force balance solutions (null space method) and rigidity analysis is shown for the same configurations. Contact forces $f_c > \langle f_c \rangle $ are mostly concentrated on the particles that belong to the over-constrained regions and there is a strong spatial correlation between the force network and the rigid network.}
\end{figure*}

\vspace{-2mm} 
In Fig. \ref{fig9}, we show the evolution of rigid clusters and the force network as a function of strain, for one initial configuration strained using the AQS protocol at $\phi=0.82$. The snapshots clearly show that there is a strong spatial correlation between the rigid cluster (red network, left panel) and the force network (right panel). Particularly, contacts with forces $f_c > \langle f_c \rangle$ are concentrated on the discs that belong to the over-constrained (or stressed) regions, the sites of which are marked with blue dots. Observe that the average contact force increases with increase in the number of over-constrained sites.  Above the shear jamming transition, almost all the particles, except a few rattlers, are over-constrained and the system can have multiple force balance solutions for the same contact network. Hence the comparison becomes less revealing at larger strains. 

\begin{table}[b]
\caption{Table of small force exponents ($\theta$, $\theta_b$ ) and $g(r)$ power law exponents ($\gamma_g$). 
The stability criteria require $\theta > \frac{1}{\gamma_g} - 2 $, $\theta_b > 1 - 2 \gamma_g$. The right hand sides of the inequalities are also shown. }
\begin{tabular}{|l|l|l|l|l|l|}
\hline
 $\phi$ & $\gamma_g$ & $\frac{1}{\gamma_g} - 2$ & $\theta$  & $1 - 2\gamma_g$ & $\theta_b$  \\ [1ex]
 \hline
 $0.55$ & $0.4212$ & $0.374$ & -- &  $0.1576$ & -- \\ [1ex]
 \hline
 $0.56$ & $0.4232$  & $0.363 $ & $0.404$ &  $0.1536$   & -- \\ [1ex]
 \hline
 $0.57$ & $0.4288$ & $0.332$ & $0.925$  & $0.1424$  & $0.366$ \\ [1ex]
 \hline
$0.58$ & $0.4341$ & $0.304$ & $1.199$  & $0.1328$  & $0.435$  \\ [1ex]
 \hline
$0.59$ & $0.443$ & $0.2573$ & $1.3654$  & $0.114$  & $0.571$ \\ [1ex]
 \hline
$0.61$ & $0.4574$ & $0.1863$ & $1.7242$   & $0.0852$  & $0.773$ \\ [1ex]
 \hline
$0.627$ & $0.46335$ & $0.158$ & $1.9782$  & $0.0733$  & $1.164$ \\ [1ex]
\hline
\end{tabular}
\end{table}
\pagebreak
\clearpage

\end{document}